\documentclass[12pt]{article}
\usepackage{graphicx} 
\usepackage[utf8]{inputenc}
\usepackage{geometry}
\usepackage{amsfonts}
\usepackage{booktabs}
\usepackage{amssymb}
\usepackage{amsmath}
\usepackage{lscape}
\usepackage{bbm}
\usepackage{xcolor}
\usepackage{url}
\usepackage{footmisc}
\newtheorem{theorem}{Theorem}

\newtheorem{claim}{Claim}

\newtheorem{definition}{Definition}
\newtheorem{example}[theorem]{Example}

\newtheorem{lemma}{Lemma}

\newtheorem{proposition}{Proposition}
\newtheorem{remark}{Remark}

\newtheorem{assumption}{Assumption}
\usepackage{setspace}
\usepackage{footmisc}
\usepackage[round]{natbib}
\usepackage[final]{hyperref} 

\usepackage{enumerate}   
\usepackage{tikz}
\DeclareMathOperator*{\cart}{\times}
\usepackage{accents}

\usepackage{chngcntr}
\usepackage{apptools}
\AtAppendix{\counterwithin{lemma}{section}}
\AtAppendix{\counterwithin{proposition}{section}}
\title{The Disagreement Dividend\thanks{I am grateful to Jonathan Bonham,  Renee Bowen, Simone Galperti, Nicola Gennaioli, Aram Grigoryan, Navin Kartik, Denis Shishkin, Joel Sobel, Pietro Spini, Guido Tabellini, Davide Viviano, and UCSD TBE seminar participants for helpful comments and suggestions.}}
\author{Giampaolo Bonomi\thanks{Department of Economics, UC San Diego. Email: gbonomi@ucsd.edu}}
\date{June 2024}

\begin{document}

\maketitle
\vspace{.5cm}
\begin{abstract}
We study how disagreement influences team performance in a dynamic game with positive production externalities. Players can hold different views about the productivity of the available production technologies. This disagreement results in different technology and effort choices --- ``optimistic'' views induce higher effort than ``skeptical'' views. Views are changed when falsified by evidence. With a single technology available, optimists exert more effort early on if the team also includes skeptics. With sufficiently strong externalities, a disagreeing team produces, on average, more than any like-minded team. With multiple technologies, disagreement over which technology works best motivates everyone to exert more effort. 
\vspace{.4cm}

\noindent \textit{JEL} Codes: D20, D83, D90, M11, M14. 

\noindent \textbf{Keywords:} disagreement; diversity; persuasion; teamwork; paradigm. 
\end{abstract}
\newpage
\section{Introduction}
\begin{quote}
\begin{singlespace}
\textit{``I don't feel that an atmosphere of debate and total disagreement and argument is such a bad thing. It makes for a vital and alive field.''
\begin{flushright} (Clifford Geertz)\end{flushright}}  
\end{singlespace}
\end{quote}

Conventional wisdom suggests that the interaction of people with different backgrounds and perspectives will often lead to socially desirable outcomes. Indeed, economists have advanced intuitive arguments linking diversity to successful problem-solving and improved decision-making, relying on the idea that different perspectives and capabilities naturally serve as complements, enriching and refining each other \citep[e.g.,][]{Hong2001,Hong2004,Page}. There is much less focus on the productivity implications of open and persistent \textit{disagreement}, characterized by strongly conflicting views and interpretations of a problem. In this paper we address the following question: can disagreement in a team of innovators motivate effort and boost the team's output? When should we expect this to happen? We focus on a specific force that characterizes disagreement: the incentive to persuade others through production breakthroughs.

Seminal contributions to the economic literature have warned us about the perils of preference disagreement, shown to create impasse and inefficiencies in many domains of social interaction.\footnote{For instance, disagreement has been shown to impede decision-making and compromise economic outcomes in social choice \citep[e.g.][]{Arrow1951}, communication \citep[e.g.,][]{CS}, public finance and public good provision \citep[e.g,][]{AlesinaT1990,AlesinaLaF2005}.} Yet, history suggests a link between the conflict of worldviews and greater innovation. The development of the first iPhone was, anecdotally, a story of disagreement \citep{iPhone,Grant2021}. Steve Jobs initially thought the product would only appeal to a ``pocket protector'' crowd, and saw the project as a dead end. A team of hard-working engineers --- with Apple's design chief Jonathan Ive on their side ---  had a very different opinion. They believed that the touchscreen technology would represent a paradigm shift for the industry. Job's skepticism meant that the team needed to design a prototype so good that it would have been impossible for him not to change his mind.\footnote{According to \citet{Grant2021}: \textit{``Fadell and his engineers chipped away at the resistance by building early prototypes in secret, showing Jobs demos, and refining their designs.''}} Some argue that Job's open disagreement culture was the key to the company's success in those years \citep[e.g.,][]{Scott2017,Grant2021}.    

Beyond the iPhone anecdote, the power of disagreement has been recognized in many innovation-related contexts. Open disagreement between scientists has led to the production of more and better theories, resulting in the belief that scientific skepticism --- the tendency to challenge and falsify existing theories --- lies naturally at the heart of scientific progress \citep{Kuhn1962}. Peers' skepticism has typically motivated philosophers to design sophisticated arguments in favor of their worldview, to convince others to adopt it.\footnote{For instance, determined to convince skeptics about the existence of God, Anselm of Canterbury designed the first \textit{onthological argument}, a class of arguments that has fascinated philosophers for almost a thousand years (see \url{https://plato.stanford.edu/entries/ontological-arguments/}).} Artists have often found in partners' disagreement and competition of ideas a motivating force inspiring them to innovate and often reach  success. \citet{Metallica} illustrate this point using the case study of the heavy metal band \textit{Metallica} and and many of their historical producers (most notably, producer Bob Rock): \textit{``Both sides [Metallica and Rock] wanted to make the best record in the world but disagreed on what it
should be like and how it should be done. [...] the most constructive outcomes of conflicts realized when Metallica collaborated
with its partners but at the same time also competed with them. The partners first competed
to find the best idea, after which the best idea was further developed together.''} 

Outside the realm of intellectual and artistic debates, the \textit{space race} was perhaps one of the starkest examples of how the competition of conflicting models (of production and society) spurred technological investment, leading to remarkable innovations in an attempt to prove the superiority of one worldview over the other \citep{TechPolitik}.\footnote{According to Brian C. Odom, Nasa Chief Historian: \textit{``In the global South, you had a lot of countries becoming independent from former colonial powers. What system would they follow? Would they follow the U.S. liberal democracy or would they follow the Soviet example of communism? Kennedy saw the race to the moon as a way to demonstrate American technological power and the benefit of one system over another.''} See \url{https://www.space.com/space-race.html}).} The more recent \textit{standards wars} in the Tech industry --- battles for market dominance between incompatible technologies --- can be thought of in a similar fashion.\footnote{R\&D breakthroughs are recognized as typical ways of persuading the key players and winning such wars. This was the case, for instance, in the battle between the AC and DC technologies for the genetation and distribution of power. See \citet{ShapiroVarian} for a discussion of  standard wars. }

The above examples share three common features: (i) the players involved hold extremely different views about one or more production technologies and are aware of such disagreement; (ii) one or more parties benefit from making others adopt their views; and (iii) persuasion occurs through the successful results of productive effort. This paper explores the productivity implications of disagreement in a simple class of games sharing the above three characteristics. Two agents engage in a two-period production game with positive production externalities. In each period, they simultaneously choose levels and allocation of costly effort across different production technologies.  At the end of each period, each player's output is realized, and payoffs materialize. 

Our set-up has the following distinctive characteristics. First, players can hold different models of the stochastic process generating returns on effort. A model defines players' view (optimistic or skeptical) about the productivity of each technology and different models imply different optimal technology and effort choices. If players have different models, they \textit{disagree} and think that the other player's model is misspecified.
Second, after observing output realizations, each player questions her model via a likelihood ratio test: she switches to the model of the other player if surprised by evidence that only such model can easily rationalize.
Finally, we assume that alternative technology views are such that, when one view is correct, higher effort in a technology conveys more information about its true productivity.

We find that, when there is only one production technology, disagreement over its productivity motivates the optimistic player to exert more effort early on. The optimist works harder in the first period, to obtain early breakthroughs that prove to the skeptic that the technology is worth the effort. In contrast, the skeptic works weakly less when paired with an optimist, to keep the latter from turning skeptic and decreasing future effort. 

We provide intuitive conditions that make the upward pressure on effort prevail, and propose a simple application to team formation: an output-maximizing manager who needs to form two teams from a pool of two optimists and two skeptics might benefit from pairing together co-workers who disagree. Following a similar logic, we show that --- when production externalities are sufficiently strong --- adding a skeptic to a team of optimists increases output more than adding an additional optimist, even if optimists always work more than skeptics. 

We build on the main insights of the single-technology case to study the implications of disagreement when alternative production technologies are available. When two technologies are similarly effective, a group of players who disagree over which technology works best is more productive than any group of individuals who share the same model --- even if the latter believe that all the technologies are highly productive. The same holds if we conjecture that one technology must be better than the other, but both seem equally promising ex-ante. By working harder in the first period, each player expects to prove --- through her breakthroughs --- the superiority of her production technology, convincing others to adopt it. As a result, more breakthroughs are obtained early on.

Our results rely on a key feature of the disagreeing technology views: according to both the skeptic and the optimistic view, the more effort is invested in a technology, the easier it becomes to tell which view correctly describes such technology. This requires some restrictions on the alternative views considered,\footnote{It does not, however, constrain the \textit{objective} stochastic process mapping effort in distributions of output. In fact, many of our results hold even if both technology views are ``wrong.''} that we formalize relying on the concept of effort's Blackwell informativeness \citep{Blackwell1954}.\footnote{The constraint is satisfied when players model technologies along the lines of many effort-decision problems in the literature \citep[e.g.,][]{Keller2005,Heidues2018,Dong2018,Ba2022}.} In the conclusive sections of the paper, we discuss the implications of changing the relationship between effort and information arrival and provide examples of pairs of technology views that yield more distinctive predictions at low effort levels. We highlight how some of our main results about the optimality of disagreeing teams change in these different economic environments, and show that many of these results are restored under additional easily interpretable changes in economic primitives. 

All in all, our analysis suggests that disagreement could be beneficial if parties have an incentive to persuade each other, and the persuasion technology is productive. The first of the two conditions fails under negative production externalities, when each player is, ceteris paribus, better off if others are not successful. Hence, an additional insight, discussed in the conclusion of our analysis: as in the iPhone and Metallica examples, disagreement should be more productive if parties have a strong interest in spreading their own model (improving each others' future decisions), rather than exploiting each others' misconceptions.

\subsection{Literature Review}\label{Lit}

To reach a more structured understanding of the economic implications of imperfect models of the world and their evolution, economic theorists have recently developed foundations to incorporate model misspecification \citep[e.g.,][]{Esponda2016, Heidues2018, Rees-Jones} and ``paradigm shifts'' \citep[e.g.][]{Hong2007, Ortoleva2012} in games and decision theoretical problems. To the current day, most of the literature has focused on long-term beliefs \citep[e.g.][]{Esponda2016,Heidues2018} and long-term misspecification robustness \citep{Fudenberg,Ba2022}, while fewer papers have addressed the incentives arising when players think strategically about each others' model differences and change. \citet{Galperti2019} shows that a persuader can exploit events that are ruled out by the model of the receiver, to expand the scope of persuasion. \citet{Schwartzstein2021} characterize the scope of ``model persuasion,'' where the sender can change the receiver's mind by proposing new models that better fit past evidence. The present paper contributes to this stream of literature by addressing a related but different question: can the persuasion incentives stemming from model differences be leveraged to increase a team's output?

Some of our ideas are related to the literature on Bayesian learning with different priors \citep[e.g.,][]{Hart2020,Kartik2021}, and its applications to principal-agent problems of evidence collection \citep[e.g.,][]{Che2009,VanDerSteen} and delegation \citep[e.g.,][]{Hirsch2016}. In particular, \citet{Che2009} and \citet{VanDerSteen} show that disagreement with the principal can motivate an expert with aligned preferences to collect more evidence.\footnote{\citet{VanDerSteen} also finds that disagreement between a manager and an employee tends to reduce delegation, hinder motivation, and lower satisfaction. By focusing on different strategic incentives, we provide a complementary, more optimistic, picture of the effects of disagreement on team effort and production.} We complement this literature with an analysis of team formation and teamwork, where agents persuade each other through successful production histories. On the one hand, we find that an output-maximizing manager could benefit from pairing employees with different views, even if she knows the characteristics of the production technology and is not driven by information motives. On the other hand, our analysis suggests that the persuasion incentives of disagreeing co-workers might reduce the average output of skeptical team members. We also show that the implications of disagreement change drastically when team members' output predictions differ the most at low effort levels, as well as in the presence of negative production externalities.\footnote{As discussed in section \ref{discussion}, a shared belief that hard work obfuscates, rather than bringing to light, information about the technology promise would lower the productivity of disagreeing teams.} 

At a high level, our paper is related to the economic literature on team diversity \citep[e.g.,][]{Hong2001,Hong2004,Page,Dong2022}. We highlight how the interaction between co-workers' strategic incentives and their perspectives on the production process can define whether team members with different views will, on average, outperform homogeneous teams. Our application of the single-technology case to teamwork shares qualitative similarities with the literature on exponential bandits \citep[e.g.,]{Keller2005}, and in particular \citet{Dong2018}. However, our driving mechanism is substantially different and leads to different behavioral and welfare implications.\footnote{\citet{Dong2018} finds that asymmetric information about the risky arm can increase experimentation in a team. Her results are driven by informational asymmetries and signaling instead of open disagreement and persuasion. In \citet{Dong2018}, the player with bad news increases effort in order to pool with the player with good news. In our model, the optimistic player takes the lead, working harder bring the skeptic on board.} 

Finally, our analysis owes to seminal contributions from the literature on hypothesis testing and experiments' informativeness \citep{Blackwell1954,NeymanPearson}. 

The paper is organized as follows. In section \ref{illustration} we provide a stylized example. In section \ref{setup} we outline the general model. In section \ref{onetech} we analyze the single production technology case and in section \ref{twotech} we cover the multiple technology case. In section \ref{discussion} we discuss the robustness to changes of our key assumptions. Finally, in section \ref{exit} we draw the conclusions.

\section{A Simple Illustration}\label{illustration}

Ann, Bob, Tom, and Lea are four engineers working in the \(R\& D\) division of a tech company. One day, Ann reaches out to her three colleagues with an idea for a new product, which --- she claims --- could be a groundbreaking innovation and earn their company great technological and reputational advantages. Tom is convinced by Ann's pitch. Bob and Lea, however, are skeptical. They think that the project poses structural issues and is not worth the effort.

The players' views are summarized in table \ref{tabview}. 
  \begin{table}[h!]
  \caption{Player Views}\label{tabview}
\centering\begin{tabular}{cc|c|c|}
      & \multicolumn{1}{c}{} & \multicolumn{2}{c}{}\\
      & \multicolumn{1}{c}{} & \multicolumn{1}{c}{\(R\)}  & \multicolumn{1}{c}{$0$} \\\cline{3-4}
   & \(\mathcal{H}\) & $\quad e\quad$ & $1-e$ \\\cline{3-4}
      &\(\mathcal{L}\) & $0$ & $1$ \\\cline{3-4}
    \end{tabular}
  \end{table}
  
Ann and Tom hold view \(\mathcal{H}\), and believe that any effort \(e\in [0,1]\) in the development of a prototype will produce a breakthrough of value \(R>0\) with probability \(e\) and no breakthrough (with value 0) with remaining probability. Bob and Lea hold the skeptical view \(\mathcal{L}\), according to which any effort spent on the project is wasted. As analysts, we conjecture that Ann is right.\footnote{The following intuitions still hold if we assign strictly positive probability to both views being correct.}

Research occurs in two-member teams. Individual effort \(e\) costs \(\frac{c}{2}e^2\), \(c>0\),\footnote{Throughout the example, assume \(c>2R\), so that effort boundaries do not affect our argument.} and engineer \(i\)'s utility given effort and output is 
\begin{equation*}
    U^i(y^i,y^{-i},e^i) = y^i + \beta y^{-i} - c\frac{(e^i)^2}{2},
\end{equation*} 
where \(y^i\) and \(y^{-i}\) are the output of players \(i\) and her co-worker respectively, and \(\beta>0\) implies a positive externality from production, a reduced form to capture complementarities between players' work or motivational benefits obtaining when fellow team members get engaged in the collaboration and obtain breakthroughs. Finally, we assume that research will take place over two periods: every period, each of the two engineers chooses an effort level \(e\in[0,1]\), observes the (independently drawn) breakthroughs of both engineers and receives a payoff.

Which engineer should work with Ann if we aim at maximizing the expected output of the two-member team? The optimist Tom or the skeptic Bob, who disagrees with Ann? The answer to the question is very simple if co-workers stick to their initial views throughout the game. A player with view \(\mathcal{H}\) chooses, in every period, an effort choice \(e^\mathcal{H}=\frac{R}{c}\). In contrast, a skeptic engineer will choose \(e^\mathcal{L}=0<e^\mathcal{H}\). The two-period expected output of the team composed of Ann and Tom is greater than that of Ann and Bob, as in the latter team only Ann will put any effort into the development of the prototype. Like-minded optimists will on average innovate more than a disagreeing team.

Now, imagine that engineers can change their minds if proven wrong: each of them will adopt the view of the co-worker if and only if they observe production outcomes possible only under the latter's model. In the second period, players with models \(\mathcal{H}\) and \(\mathcal{L}\) will still pick \(e^\mathcal{H}\) and \(e^\mathcal{L}\) respectively. However, the FOC for Ann's first-period effort becomes 
\begin{equation}\label{exintro}
\underbrace{ce^\star}_{\substack{\text{marginal cost} \\ \text{of effort}}} = \underbrace{R}_{\substack{\text{marginal expected} \\ \text{static return on effort} }} + \underbrace{\beta R(e^{\mathcal{H}}-e^{\mathcal{L}})}_{\substack{\text{marginal benefit from} \\ \text{changing Bob's mind}}}.
\end{equation}
 Ann's marginal benefit from effort is now greater than \(R\): if she obtains an early breakthrough she will convince Bob to embrace view \(\mathcal{H}\) and increase effort from \(e^{\mathcal{L}}\) to \(e^\mathcal{H}\) in the future (desirable because \(\beta>0\)). As a result, her first-period effort is \(e^\mathcal{H}(1+\beta e^\mathcal{H})\), above \(e^{\mathcal{H}}\), the level of effort that she exerts in the same period if paired with Tom. If we need to form two teams working on Ann's idea (each team playing a separate production game) the joint output of the teams will be maximized by pairing Ann with Bob, and Tom with Lea. This happens for two reasons: first, from condition \ref{exintro}, Ann and Tom work harder in the first period when paired with a skeptic. Second, with probability \(e^\mathcal{H}(1+\beta e^\mathcal{H})\) an optimist obtains a breakthrough in the first period, persuading their co-worker to pick effort \(e^{\mathcal{H}}\) in the second period. Such a change of mind is not possible in like-minded teams.

Will Ann and Bob produce more than Ann and Tom? Ann exerts more effort when paired with Bob but Bob always works less than Tom because he is skeptical. It is easily shown that the former force prevails if \(\beta>\frac{2-\frac{R}{c}}{\frac{R}{c}(1+\frac{R}{c})}\). With sufficiently strong externalities Bob's skepticism will motivate Ann to work very hard, increasing the team's overall output above the performance of two like-minded optimists.

Imagine now that Bob proposes an alternative approach for the development of the innovative product. Let \(x\) denote Ann's approach, and \(y\) Bob's proposal. Bob is optimistic about \(y\), but Ann is skeptical about it. Tom is optimistic about both ideas and --- after a change in mind --- Lea is too. The views of the four engineers can be summarized by four tuples, respectively \((\mathcal{L}_x,\mathcal{H}_y), (\mathcal{H}_x,\mathcal{L}_y), (\mathcal{H}_x,\mathcal{H}_y)\) and \((\mathcal{H}_x,\mathcal{H}_y)\), which we call \textit{models}. In every period, each engineer will choose (i) which approach to work on; and (ii) how much research effort to exert. Between periods, she adopts her co-worker's model if first-period returns have zero probability based on her initial model but not the co-worker's. 

If \((\mathcal{H}_x,\mathcal{H}_y)\) is correct, which team will on average innovate more, the opposite-minded Ann and Bob, or the like-minded optimists, Tom and Lea?

Tom and Lea will remain like-minded and choose \(e^{\mathcal{H}}\) in every period. Whether or not they still disagree, Bob and Ann will also put effort \(e^\mathcal{H}\) in the second period, choosing the approach about which they are optimistic.\footnote{It is easily verified that, in the second period, each engineer will operate a technology for which she is optimistic, provided that she enters the period optimistic about some technology.} However, in the first period, Ann and Bob will work harder than Tom and Lea. We now illustrate why. Let \(i = A, B\) indicate Ann and Bob respectively and \(R^i_k\) be the expected marginal static return on effort of approach \(k = x,y\) according to \(i\)'s views, so that \(R^A_x = R^B_y = R\) and \(R^A_y = R^B_x = 0\). In the first period, Ann will choose to work on \(x\) and Bob on \(y\).\footnote{Ann believes that using \(y\) exhibits the following disadvantages relative to \(x\): (i) it will never lead to any breakthrough; and (ii) it will never convince Bob to change model. Both are undesirable.} From the FOC, Ann's first-period effort \(e^A\) satisties
\begin{equation}\label{exintro2}
\underbrace{ce^A}_{\substack{\text{marginal cost} \\ \text{of effort}}} = \underbrace{R}_{\substack{\text{marginal expected} \\ \text{static return on effort} }} + \underbrace{\beta(R^A_x - R^A_y)e^{\mathcal{H}}}_{\substack{\text{marginal benefit from making} \\ \text{Bob switch approach}}}
\end{equation}
and, similarly, for Bob,
\begin{equation}\label{exintro3}
\underbrace{ce^B}_{\substack{\text{marginal cost} \\ \text{of effort}}} = \underbrace{R}_{\substack{\text{marginal expected} \\ \text{static return on effort} }} + \underbrace{\beta(R^B_y - R^B_x)e^{\mathcal{H}}}_{\substack{\text{marginal benefit from making} \\ \text{Ann switch approach}}}.
\end{equation}
The second term on the right-hand side of \ref{exintro2} has the following interpretation. Ann thinks that Bob's initial model makes him wastefully allocate effort \(e^\mathcal{H}\) to the bad approach \(y\). Since \(\beta>0\), she subjectively benefits from convincing him to switch to \((\mathcal{H}_x, \mathcal{L}_y)\), as the change in mind will cause Bob to adopt approach \(x\) in the second period, granting Ann an expected utility gain of \(\beta(R^A_x - R^A_y)e^{\mathcal{H}}\). Clearly, the same reasoning holds for Bob. From conditions \ref{exintro2} and \ref{exintro3}, one can easily see that \(e^A = e^B = e^\mathcal{H}(1+\beta e^\mathcal{H})>e^{\mathcal{H}}\). Note that we also expect both Ann and Bob to earn higher payoffs, throughout the game, than Tom or Lea as disagreement pushes players' first-period effort closer to \(e^{\mathcal{H}}(1+\beta)\) the effort level that is efficient taking externalities into account.\footnote{While the main focus of our analysis will be output, we discuss welfare implications in section \ref{discussion}.}

In the remainder of the paper, we generalize, enrich, and qualify the intuitions presented in the previous examples.

\section{The Production Game} \label{setup}
This section describes the production game. After presenting the characteristics of the objective production process and players' subjective models, we introduce the notion of actions' informativeness. We then illustrate how players change models. Finally, we describe the solution concept and conclude the section by discussing our main assumptions.   

\subsection{Objective Process}

Ann (A) and Bob (B) engage in a production over two periods.\footnote{A two-period time horizon is the simplest setting that allows us to isolate the dynamic incentives at the core of our theory, which generalizes to richer dynamic settings.} In every period, they have access to a finite number of production technologies, each yielding output according to a technology-specific stochastic process. Formally, each technology \(k\in K\), \(|K|\ge 1\), is characterized by an \textit{objective} distribution, \(Q_k:\mathcal{E}\to \Delta(\mathcal{Y})\), over the set of output realizations \(\mathcal{Y}\subseteq \mathbb{R}\) for any effort level in \(\mathcal{E} = [0,b]\), \(b>0\).
 When Ann chooses to put \(e^A\in \mathcal{E}\) in technology \(k\), her output \(Y^A\) is drawn from \(Q_k(\cdot|e^A)\).  Conditional on effort choices, output is distributed independently across technologies, players, and periods, and, for simplicity, we assume that each player can operate at most one technology per period.

Player \(i = A,B\) has a stage payoff function \(U^i:\mathcal{Y}^2\times \mathcal{E}\to \mathbb{R}\) of the following form, 
\begin{equation*}
U^i(y,e^i) = u(y^i,e^i) + v(y^{-i}),
\end{equation*}
where \(y^i\) is the output obtained by player \(i\) in the stage production activity, \(y^{-i}\) is the output of the other player, \(u:\mathcal{Y}\times\mathcal{E}\to\mathbb{R}\) captures utility from own output, with \(\frac{\partial u(y^i,e^i)}{\partial y^i}>0\) and  \(\frac{\partial u(y^i,e^i)}{\partial e^i}<0\). Finally, \(v:\mathcal{Y}\to\mathbb{R}\) captures production externalities across players. We assume that \(\frac{\partial v(y^{-i})}{\partial y^{{-i}}}>0\) so that there are positive externalities from production.

For every technology choice, higher effort yields higher output realizations more often than lower effort, in the sense described by the following definition.
\begin{definition}[FOSD monotonicity]
Let \(F: \mathcal{D}\to\Delta(\mathcal{A})\), where \(\mathcal{A},\mathcal{D}\subseteq\mathbb{R}\). We say that: 
\begin{enumerate}[(i)]
\item \(F\) is FOSD-monotone in \(\mathcal{D}\) if, for every \(d,\tilde{d}\in \mathcal{D}, \tilde{d}>d\),  either \(F(\cdot|\tilde{d})\) first-order stochastic dominates\footnote{Let \(F\) and \(G\) be two probability distributions with support on the real line, and let \(A\) and \(B\) be distributed according to \(F\) and \(G\) respectively. We say that \(F\) first order stochastically dominates \(G\) if \(\mathbb{P}_{F}(A\le x)\le\mathbb{P}_{G}(B\le x)\) for every \(x\in \mathbb{R}\), and there exists \(x\in\mathbb{R}\) such that \(\mathbb{P}_{F}(A\le x)<\mathbb{P}_{G}(B\le x)\).} \(F(\cdot|d)\) or
\(F(\cdot|\tilde{d}) = F(\cdot|d)\).
\item \(F\) is strictly FOSD-monotone in \(\mathcal{D}\) if, for every \(d,\tilde{d}\in \mathcal{D},\tilde{d}>d\),  \(F(\cdot|\tilde{d})\) first-order stochastic dominates \(F(\cdot|d)\).
\end{enumerate}
\end{definition}
\begin{assumption} \label{ass1}
For each \(k\in K\), \(Q_k\) is strictly  FOSD-monotone in \(\mathcal{E}\).
\end{assumption}

\subsection{Subjective Models and Disagreement}
We assume that \(Q = (Q_k)_{k\in K}\) is not common knowledge, but players hold subjective models of the output process and can be optimistic or skeptical about each technology. Formally, for each technology \(k\), players hold one of two \textit{technology views} \(\mathcal{H}\) or \(\mathcal{L}\), with \(\mathcal{H}: \mathcal{E}\to\Delta(\mathcal{Y})\) and \(\mathcal{L}: \mathcal{E}\to\Delta(\mathcal{Y})\). Before characterizing the two views, we introduce some notation. 

For any function \(g:\mathcal{Y}\to\mathcal{Y}\), we write \(\mathbb{E}_{\mathcal{H}}[g(Y)|e,k]\) and \(\mathbb{E}_{\mathcal{L}}[g(Y)|e,k]\) to refer to a player's expectation of \(g(Y)\) when she invests effort \(e\) in technology \(k\), for which she holds view \(\mathcal{H}\) and \(\mathcal{L}\) respectively. Note that \(\mathbb{E}_{\mathcal{L}}[g(Y)|e,k] = \mathbb{E}_{\mathcal{L}}[g(Y)|e,k^\prime]\) and \(\mathbb{E}_{\mathcal{H}}[g(Y)|e,k] = \mathbb{E}_{\mathcal{H}}[g(Y)|e,k^\prime]\) for any \(k,k^\prime \in K\), so that the \(k\) index is only used to keep track of the technology adopted. 

\begin{assumption}\label{ass2}
Technology views satisfy the following properties.
\begin{enumerate}[(i)]
\item View \(\mathcal{L}\) is FOSD-monotone in \(\mathcal{E}\) and view \(\mathcal{H}\) is strictly FOSD-monotone in \(\mathcal{E}\).
\item For all \(e\in \mathcal{E}\), with \(e>0\), \(\mathcal{H}(\cdot|e)\) first order stochastically dominates \(\mathcal{L}(\cdot|e)\). In addition, either \(\mathcal{H}(\cdot|0)\) first order stochastically dominates \(\mathcal{L}(\cdot|0)\) or \(\mathcal{H}(\cdot|0) = \mathcal{L}(\cdot|0)\).
\item For all \(m_{k}\in \{\mathcal{H},\mathcal{L}\}\), \(E_{m_k}[u(Y,e)|e,k]\)   
is continuous in \(e\), differentiable in the interior of \(\mathcal{E}\), and has a unique maximizer \(e^{m_k}\in \mathcal{E}\) with \(b>e^{\mathcal{H}}>e^{\mathcal{L}}\ge 0\).
\end{enumerate}
\end{assumption}
Part (i) requires that, regardless of the view, players expect higher effort to yield high-output realizations more often, strictly so in the case of model \(\mathcal{H}\). Part (ii) implies that \(\mathcal{H}\) leads to more \textit{optimistic} output expectations than \(\mathcal{L}\) for any effort level. The main implication of (iii) is that  \(\mathcal{H}\) encourages strictly more effort than \(\mathcal{L}\) in the single decision-maker problem where the objective is maximizing expected utility from own productive activity. 

A player's subjective \textit{model} is a collection of her technology views, one for each technology.
\begin{definition}\label{def1}
A model is an element of \(M = \cart_{k\in K}\{\mathcal{H},\mathcal{L}\}\).
\end{definition}
Models are reduced-form representations of how Ann and Bob organize the consequences of their productive decisions. They pin down a distribution on output realizations for any effort and technology choices. While their models might be different from \(Q\), players correctly assume that output realizations are independent across technologies, players, and periods.

In our analysis we allow players to hold different models: \(m^A\) could be different from \(m^B\). At the same time, each player knows the model adopted by the other. This is our notion of \textit{disagreement}: players know that they are modeling the game differently, but still think their own model corresponds to the truth \(Q\) and that any other model is misspecified.

\paragraph{Effort Informativeness}
Different actions can carry different informational content about the correctness of each view, in the sense of the following definition. 

\begin{definition}\label{def2}
For any \(e,e^\prime\in \mathcal{E}\) we say that \(e^\prime\) is more informative than \(e\) if experiment\footnote{We adopt the following notation for experiments with an (arbitrary) dichotomous state space \(\Omega = \{\omega_1,\omega_2\}\). Any dichotomous experiment is characterized by the touple \((G,P,S)\), where \(S\) is a signal space, \(G\) is the distribution on \(S\) conditional on the state being \(\omega_1\), and \(P\) is the distribution on \(S\) conditional on the state being \(\omega_2\).} \(\Pi_{e^\prime} =(\mathcal{H}(\cdot|e^\prime),\mathcal{L}(\cdot|e^\prime),\mathcal{Y})\) is Blackwell more informative than \(\Pi_{e}=(\mathcal{H}(\cdot|e),\mathcal{L}(\cdot|e),\mathcal{Y})\).
\end{definition}

\begin{assumption}\label{ass3}
For each \(e^{\prime},e\in \mathcal{E}\), action \(e^\prime\) is more informative than \(e\) if and only if \(e^\prime>e\).
\end{assumption}
The interpretation of the assumption is as follows. Disagreeing players believe that hard work yields on average higher levels of output, makes breakthroughs more likely, and therefore should make it easier to evaluate the \textit{relative} fit of alternative views of the technology adopted. At the beginning of the appendix, we propose two intuitive examples that satisfy our assumptions on technology views. They closely recall bandit problems \citep{Keller2005} as well as the effort decision model proposed by \citet{Heidues2018}.

\subsection{Game Timeline}
\begin{figure}[h!]
    \centering
    \includegraphics[width=.9\textwidth]{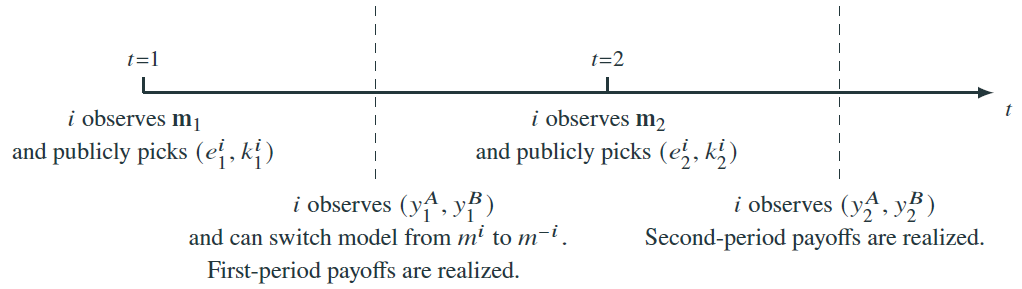}
    \caption{Timeline} \label{fig1}
\end{figure}

At the beginning of period \(t=1,2\),  Ann and Bob simultaneously choose their production technologies \(k_t^A\) and \(k_t^B\), and effort levels \(e_t^A\) and \(e_t^B\), respectively. Actions \(\mathbf{k}_t = (k_t^A,k_t^B)\) and \(\mathbf{e}_t = (e_t^A,e_t^B)\) are publicly observed. At the end of the period, output levels \(y_t^A\) and \(y_t^B\) --- grouped in profile \(\mathbf{y}_t\) --- are drawn from \(Q_{k_t^A}(\cdot|e^A_t)\) and \(Q_{k_t^B}(\cdot|e^B_t)\) and publicly observed. Stage payoffs are realized.
The game timeline is reported in figure \ref{fig1}. 

 Ann and Bob start the game with models \(m^A_1\) and \(m^B_1\) respectively, which are common knowledge and grouped in profile \(\mathbf{m}_1\). At the end of \(t=1\), models evolve following rule \ref{LR} presented in the next section, which determines the models \(\mathbf{m}_2 = (m^A_2,m^B_2)\) held in \(t=2\). In what follows, we drop time subscripts if it does not give rise to ambiguity.

\subsection{Model Change}
Ann and Bob are persuaded to change their model if they face surprising evidence that falsifies it in favor of the model of their co-worker. In particular, first-period actions and production outcomes \((\mathbf{e},\mathbf{k},\mathbf{y}) = (e^A,e^B,k^A,k^B,y^A,y^B)\) are publicly observed by the end \(t=1\), and player \(i = A,B\) will switch (not switch) model from \(m^i\) to \(m^{-i}\) if 
\begin{equation}\label{LR}
L_{m^{-i}}(\mathbf{y}|\mathbf{e},\mathbf{k}) > (<)\medspace c^\alpha_{m^i}(\mathbf{e},\mathbf{k}, \mathbf{m}) L_{m^i}(\mathbf{y}|\mathbf{e},\mathbf{k})\tag{LR}
\end{equation}
where \(L_{m^{i}}\) represents the likelihood of the observed production outcomes according to model \(m^i\); \(\mathbf{m} = (m^A,m^B)\) is the profile of first period models of Ann and Bob; and \(c^\alpha_{m^i}\) is defined as the smallest non-negative scalar such that the probability of switching away from model \(m^i\) if \(Q=m^i\), i.e. the type I error, is at most \(\alpha\).\footnote{Following a standard practice in the construction of likelihood ratio tests, if there is equality between the right-hand side and left-hand side of relation \ref{LR}, player \(i\) will switch with probability \(q\), chosen so that the type I error probability is exactly \(\alpha\)  (the randomization outcome being independent across players, and independent from the production history). For convention, we assume that in case of output realizations outside the support of both \(m^A_1\) and \(m^B_1\), no change occurs.}  The coefficient \(\alpha\in [0,1]\) captures the degree of flexibility of players and is assumed small, so that players tend to resist model changes.

Denote by \(\phi_{m^{i}}^{m^{-i}}(\mathbf{e},\mathbf{k}|m^i)\) the probability of a switch from \(m^{-i}\) to \(m^i\) conditional on \(Q = m^i\) and action profile \((\mathbf{e},\mathbf{k})\). We make the following assumption.

\begin{assumption}
For each \(m,m^\prime\in M\), each \((\mathbf{e},\mathbf{k})\in \mathcal{E}^2\times M^2\), and \(i= A,B\) it holds that \(\phi_{m^{i}}^{m^{-i}}(\mathbf{e},\mathbf{k}|m^i)\) is (i) continuous, and (ii)  differentiable in \(e^i\) when \(\phi_{m^{i}}^{m^{-i}}(\mathbf{e},\mathbf{k}|m^i)<1\).
\end{assumption}
Requirement (i) guarantees the existence of an equilibrium, while requirement (ii) simplifies exposition. We now describe the solution concept used throughout the analysis.

\subsection{Hypothesis Testing Equilibrium}
A stage strategy \(s^i_t = (s^i_{te},s^i_{tk})\) for player \(i\) consists of an effort rule \(s^i_{te}: M^2\to \mathcal{E}\) and a technology rule \(s^i_{tk}: M^2\to K\) specifying, respectively, \(i\)'s effort and technology choice for each possible pair of models held  by the two players at the beginning of \(t\). A strategy profile \(s^i = (s^i_1,s^i_2)\) consists of a stage strategy for each \(t\). We use the following equilibrium notion.\footnote{ To simplify exposition the analysis focuses on pure strategies, but allowing for mixed strategies does not change our results. Equilibrium existence never requires randomization in the second period. Randomization is never required in equilibrium in the first period if Ann and Bob start the game with the same model. At this level of generality, in specific instances, mixing in the first period might be required for equilibrium existence when players start the game with different models. In those cases, all our propositions still hold, and the inequalities of lemma \ref{lemmaeffort} hold for each element in the supports of Ann and Bob's mixed strategies.}

\begin{definition}[HT Equilibrium]
A profile \((s^A,s^B)\) is an HT equilibrium of the game if:
\begin{enumerate}[(i)]
\item For each profile of second-period models \(\mathbf{m}_2 \in M^2\) and each \(i\in I\), \(s_2^i(\mathbf{m}_2)\) solves
\begin{equation*}
\max_{(e,k)\in\mathcal{E}\times K}\mathbb{E}_{m^i_2}\left[U^i(Y,e)|k,s_2^{-i}(\mathbf{m}_2)\right]
\end{equation*}
\item For each profile of first-period models \(\mathbf{m}_1\in M^2\) and each \(i\in I\), \(s_1^i(\mathbf{m}_1)\) solves
\begin{equation*}
\max_{(e,k)\in\mathcal{E}\times K}\mathbb{E}_{m^i_1}\left[U^i(Y,e) + \delta V^i_{s_2,m^i_1}(\mathbf{m}_2)\Big|e,k,s_1^{-i}(\mathbf{m}_1), \mathbf{m}_1\right],
\end{equation*}
given that player \(i\) anticipates that the transition from \(\mathbf{m}_1\) to \(\mathbf{m}_2\) will follow rule \ref{LR}, \(\delta\in (0,1]\), and \(V^i_{s_2,m^i_1}(\mathbf{m}_2)\) is defined as follows, 
\begin{equation*}
V^i_{s_2,m^i_1}(\mathbf{m}_2) = \mathbb{E}_{m^i_1}\left[U^i(Y,s^i_{2e}(\mathbf{m}_2))|s^i_{2k}(\mathbf{m}_2),s_2^{-i}(\mathbf{m}_2)\right].
\end{equation*}
\end{enumerate}
\end{definition}
Without loss, we set \(\delta = 1\). Note that players are aware of disagreement and account for how their behavior in the first period might affect the future value of the game through model changes. Yet, they dogmatically assume that their own current model is correct. 

It remains to define the criterion used to compare the performance of different teams. Let \(\hat{S}\) be the set of HT equilibria of the game. For each \(s\in \hat{S}\) and \textit{initial} models \(m^A,m^B\in M\) let
\(Y_s(m^A,m^B,Q) = \sum_{i\in \{A,B\}}\sum_{t\in\{1,2\}}\mathbb{E}_Q\left[y^i_t|m^A,m^B,s\right]\),
where the expectation is based on \(Q\) and assumes that players play according to \(s\) with initial models \(\mathbf{m_1} = (m^A,m^B)\). Define
\begin{equation*}
\hat{Y}(m^A,m^B,Q) = max_{s\in \hat{S}}Y_s(m^A,m^B,Q),
\end{equation*}
so that, for instance, \(\hat{Y}(m^A,m^B,Q)>\hat{Y}(m^A,m^A,Q)\) means that the expected output of the ``most productive'' equilibrium is higher when Ann and Bob disagree than when they share Ann's model.\footnote{While the focus on the most productive equilibria provides a simple measure of team performance when there are multiple equilibria, the results of the following sections hold for every equilibrium.} Note that, for each \(m,m^\prime\in M\), \(\hat{Y}(m^\prime,m,Q)=\hat{Y}(m,m^\prime,Q)\), by symmetry.

\subsection{Discussion of the Main Assumptions}\label{AssDisc}
\paragraph{Payoffs} The additive separability of the utility function isolates the persuasion incentive created by disagreement as the only source of strategic interaction between Ann and Bob. This allows us to better highlight the driving force of our results.\footnote{For the same reason, the output stochastic process rules out technological complementarities between players' efforts. \citet{Prat2002} studies the relation between such complementarities and team heterogeneity.} Positive externalities are a reduced-form way of capturing how the overall team performance in a project can feedback positively on each team member. The sign of externalities plays a fundamental role in our analysis and in section \ref{discussion} we discuss how our results change under negative externalities.

\paragraph{Action Informativeness} Assumption \ref{ass3} is satisfied when technology views take the form of many known effort decision models \citep[e.g.,][]{Keller2005,Heidues2018,Dong2018,Ba2022}. Naturally, in bandit problems higher effort increases the probability of learning the quality of an arm, but the property extends to other applications (e.g., example \ref{ex2} in the appendix). If the relation between effort and information was negative rather than positive, our results would change substantially, but results are robust to changing such relation and the sign of externalities simultaneously. We illustrate this point in section \ref{discussion}.

\paragraph{Disagreement and Testing} Players can agree to disagree over the true model. In this case, Ann sees Bob's model as misspecified and uninformative about \(Q\). The test \ref{LR} implies that players ignore evidence against their model, unless surprised.\footnote{The likelihood ratio test was popularized by \citep{NeymanPearson}. Variations of this rule have been adopted in many forms in the economic literature \citep[e.g.,][]{Hong2007,Ortoleva2012,Ba2022}.} This captures two largely documented behavioral biases: overconfidence in own worldviews \citep[e.g.,][for a review]{Evans1990,Nickerson1998,Andreoni2012,Galperti2019}, and over-reaction to low probability events \citep[for a review,][]{Hong2007, Ortoleva2012}. Finally, players only consider alternative models held by other group members. If they disagree, this captures disagreement-induced binary thinking \citep{Lewis}. If they are like-minded, it captures groupthink, inhibiting their ability to change perspectives \citep{Janis82}.\footnote{Our qualitative results could be replicated in a Bayesian framework under additional assumptions. However, our driving force would be harder to isolate due to players' own learning incentives. Most importantly, we believe that the model change approach of this paper better matches our motivating examples, while also qualitatively capturing behavioral biases shown to arise when different worldviews collide.} 

\section{One Technology: The Power of Skepticism} \label{onetech}

We start our analysis from the case of a single technology, \(|K|= 1\). In this case, we drop the technology subscript \(k\), as strategies only consist of effort rules, and technology views and models coincide. We refer to the agent with model \(\mathcal{H}\) as the \textit{optimistic} player, while the agent with model \(\mathcal{L}\) is the \textit{skeptic}. The question that underlies the section is the following: when is a disagreeing team expected to produce more (or less) than like-minded teams?

We start by making two observations. First, due to the separability of stage payoffs, in the second period a player with model \(\mathcal{H}\) (or \(\mathcal{L}\)) exerts the same amount of effort \(e^\mathcal{H}\) (or \(e^\mathcal{L}\)) regardless of the model of the opponent. By assumption \ref{ass2}, \(e^\mathcal{H}>e^\mathcal{L}\). Second, when one of the technology views is correct, higher effort makes a switch towards such view more likely.

\begin{lemma}\label{lemmapower}
For all \(m^A,m^B\in M\) and \(i\in I\), \(\phi_{m^{i}}^{m^{-i}}(\cdot,\cdot|m^i)\) is increasing in \(e^A\) and \(e^B\).
\end{lemma}
The lemma is proven in the appendix. It has a simple interpretation: both Ann and Bob believe that by working harder they will prove to their disagreeing counterpart that they hold the true model \(m^i\). This is always true \textit{subjectively}, regardless of the true model \(Q\).\footnote{The result follows from our assumption on effort informativeness and known results by \citet{NeymanPearson} and \citet{Blackwell1954}} The result is at the core of the persuasion incentives highlighted in the paper: players believe that more productive actions --- higher effort --- will be more likely to trigger a model change. This force is akin to the ``information validates the prior'' finding of \citet{Kartik2021} in a Bayesian setting.\footnote{\citet{Kartik2021} shows that, under some ordering conditions, if Ann and Bob have different priors, Ann expects more information to move Bob's posterior mean closer to her prior mean.} 

Our next result illustrates how the incentives to facilitate or prevent a model change by the other player modify Ann and Bob's effort decisions when they disagree.

\begin{lemma}\label{lemmaeffort}
Consider any equilibrium \(s\) of the game with \(|K|=1\) and let the initial models be \(m^A = \mathcal{H}\) and \(m^B = \mathcal{L}\). The following holds about first-period effort.
\begin{enumerate}[(i)]
    \item Ann exerts more effort than when Bob is like-minded, \(s^A_{1e}(\mathcal{H},\mathcal{L})\ge s^A_{1e}(\mathcal{H},\mathcal{H})\)
    \item Bob exerts less effort than when Ann is like-minded, \(s^B_{1e}(\mathcal{H},\mathcal{L})\le s^B_{1e}(\mathcal{L},\mathcal{L})\).
\end{enumerate}
The inequality of part (i) holds strictly whenever \(\phi_{m^{i}}^{m^{-i}}(\cdot,\cdot|m^i)\) is strictly increasing in \(e^i\).
\end{lemma}
The intuition for this result is as follows. First, Ann and Bob anticipate that if they exert more effort, they will produce more information about the production technology. If they disagree, they expect this to falsify the model of the other group member more often, by lemma \ref{lemmapower}. Next, note that if Bob switches to model \(\mathcal{H}\), he will work harder in the second period, because \(e^\mathcal{H}>e^{\mathcal{L}}\). This benefits Ann through the production externality. Conversely, when Ann is the one changing her mind after the first period --- switching to \(\mathcal{L}\) --- her effort and expected output in the second period will decrease, lowering Bob's future expected payoff. As a consequence, Ann works more in the first period than she would if Bob was like-minded, in order to ``bring him on board.'' Bob, in contrast, might be induced to reduce effort to prevent Ann from becoming skeptical. However, when  \(e^{\mathcal{L}} = 0\), as in section \ref{illustration}, Bob's effort does not decrease under disagreement, as he's not engaging in production in the first place.

The rest of the section presents implications of lemma \ref{lemmaeffort}. We start by comparing the performance of a disagreeing team to the one of the average like-minded team. 

\begin{proposition}\label{prop2formation}
Assume that \(\mathcal{L}\) discourages effort, \(e^{\mathcal{L}}=0\). Then, if the optimist holds the true model, a disagreeing team is expected to produce more than the average of the two like-minded teams. If the skeptic holds the true model, a disagreeing team can be expected to produce more or less than the average of the two like-minded teams. 
\begin{gather*}
2\hat{Y}(\mathcal{H},\mathcal{L},\mathcal{H}) \ge \hat{Y}(\mathcal{H},\mathcal{H},\mathcal{H}) + \hat{Y}(\mathcal{L},\mathcal{L},\mathcal{H}) \\
2\hat{Y}(\mathcal{H},\mathcal{L},\mathcal{L}) \gtreqless  \hat{Y}(\mathcal{H},\mathcal{H},\mathcal{L}) + \hat{Y}(\mathcal{L},\mathcal{L},\mathcal{L}).
\end{gather*}
The first inequality holds strictly if \(\phi_{m^{i}}^{m^{-i}}(\cdot,\cdot|m^i)\) is strictly increasing in \(e^i\). 
\end{proposition}

In the appendix, we propose a corresponding set of results, collected in proposition \ref{prop1bench}, where we relax the assumptions that players are forward-looking and aware of disagreement. By comparing these results, the reader can verify how the implications of lemma \ref{lemmaeffort}, which rely on both awareness of disagreement and dynamic incentives, play a key role in our results.

Proposition \ref{prop2formation} tells us that when the true technology is \(Q = \mathcal{H}\) and \(\mathcal{L}\) discourages effort (\(e^{\mathcal{L}} = 0\)), we can expect a disagreeing group to produce more than the average like-minded group. A comparison with proposition \ref{prop1bench} shows that this disagreement premium is absent if players are unaware of disagreement and smaller if they are myopic.  Next, when \(Q = \mathcal{L}\) disagreement could both increase and decrease aggregate expected output. The possibility that output goes up owes entirely to the persuasion incentives of the optimistic player, which push first-period output upwards. It is therefore absent in \ref{prop1bench}.

The intuition for proposition \ref{prop2formation} is the following. When skeptics opt out from production, two forces are in place in the disagreeing group. First, the optimistic player works harder in the first period, increasing aggregate average production in \(t= 1\). Second, more information arrives about the underlying true model, due to higher effort in period \(t=1\). This second force will drive second-period output up if the true model is \(Q = \mathcal{H}\) and down if \(Q = \mathcal{L}\), since it makes a switch towards the true model relatively more likely. The two forces increase output unequivocally if \(Q=\mathcal{H}\), but have an ambiguous total effect if \(Q=\mathcal{L}\). 

The results have the following implications for group composition. 

\paragraph{Team Formation} \textit{Two teams need to be formed starting from a pool of four workers, two workers are optimists and two are skeptics. Output realizations are not observed across teams. Proposition \ref{prop2formation} tells us that when the true structure of returns on effort is \(Q = \mathcal{H}\), then the aggregate output of the two teams will be (on average) larger when two disagreeing co-workers are paired together.}
\vspace{.60cm}

Proposition \ref{prop2formation} provides, we believe, interesting insights for team formation problems, but does not clarify whether a disagreeing group could outperform a team of like-minded optimists. As we show in proposition \ref{prop1bench} in the appendix, two optimists will produce more than any other group if players are unaware of disagreement or myopic. We now show that allowing for persuasion incentives changes such results.  

For the next proposition, assume that \(b\) (the upper bound on effort) is large enough that there exists \(\hat{e}_{Q}\in\mathcal{E}\) such that \(\mathbb{E}_Q[Y|\hat{e}_Q] = 4\mathbb{E}_Q[Y|e^{\mathcal{H}}]\). Let \(\Delta = \mathbb{E}_{\mathcal{H}}[v(Y)|e^{\mathcal{H}}] - \mathbb{E}_{\mathcal{H}}[v(Y)|e^{\mathcal{L}}]\).

\begin{proposition}\label{prop3}
A disagreeing group can sometimes produce more than a group of optimists,   
 \begin{equation*}
 \hat{Y}(\mathcal{H},\mathcal{L}, Q) \gtreqless  \hat{Y}(\mathcal{H},\mathcal{H}, Q).     
 \end{equation*}
In particular, the disagreeing group is more productive if externalities are strong enough. If \(\Delta> -\frac{\frac{\partial}{\partial e^A}\mathbb{E}_\mathcal{H}[u(Y^A,e^A)|e^A]}{\frac{\partial}{\partial e^A}\phi^{\mathcal{L}}_{\mathcal{H}}(e^A,e^B|\mathcal{H})}\) for all \(e^A\in [e^\mathcal{H},\hat{e}_{Q}]\) and  \(e^B\in [0,e^{\mathcal{L}}]\), then 
\begin{equation*}
 \hat{Y}(\mathcal{H},\mathcal{L}, Q) >  \hat{Y}(\mathcal{H},\mathcal{H}, Q). 
\end{equation*}
\end{proposition}
Note that the condition can always be satisfied by choosing \(v\) such that \(\Delta\) is sufficiently large, provided that \(\frac{\partial}{\partial e^A}\phi_{\mathcal{H}}^{\mathcal{L}}(e^A,e^B|\mathcal{H}) \ne 0\) in the specified range of effort levels. The phenomenon, already described in section \ref{illustration}, is further illustrated in example \ref{ex3} in the appendix.

Proposition \ref{prop3} suggests that adding to a team of optimists a skeptic (who doesn't exert effort unless convinced) could --- under some circumstances --- be more valuable to an output-maximizing team manager than adding an additional optimist: the presence of a skeptic allows the manager to get the most out of the more optimist team members, those already convinced that the effort will pay off. For the benefit of disagreement to materialize, the externalities must be sufficiently high. If Ann expects to strongly benefit from changing Bob's mind, she will increase her effort and first-period output considerably, possibly producing more than a group composed of individually most productive types. If such externalities are low, Ann will not care about Bob's view and behave similarly to a myopic player.

In the next section we show that with alternative and competing (similarly good) technologies, there always exists a type of disagreeing team that, under relatively weak conditions, will produce more than any like-minded one, even when externalities are moderate.

\section{Competition of Ideas} \label{twotech}
We now analyze the problem with \(|K|=2\), \(K = \{x,y\}\).\footnote{The extension of our main results to \(|K| > 2\) is trivial given our focus on two-member groups and the assumption that each group member only switches to a model adopted by someone in a previous stage of the game, as discussed in section \ref{AssDisc}. Consistently with the conditions of propositions \ref{prop4fixed} and \ref{prop5end}, an immediate extension requires that there exist \(x,y\in K, x\ne y,\) such that (i) \(Q_{x} = Q_{y}\), and (ii) \(Q_{x}(\cdot|e)\) weakly first-order stochastically dominates \(Q_{k} (\cdot|e)\) for any \(e\in\mathcal{E}\) and \(k\in K\setminus \{x,y\}\)} With two technologies, the set of models becomes \(M = \{(\mathcal{H}_{x},\mathcal{H}_{y}),(\mathcal{H}_{x},\mathcal{L}_{y}), (\mathcal{L}_{x},\mathcal{H}_{y}), (\mathcal{L}_{x},\mathcal{L}_{y})\}\), where subscripts are only used to keep track of the views corresponding to each technology. According to this notation, when Ann models technologies according to \((\mathcal{H}_x,\mathcal{L}_y)\), she holds the optimistic view about technology \(x\) and the skeptical one for technology \(y\).

The multi-technology setup admits many possible interpretations. For instance, our group could consist of board members of a company trying to decide which business project to prioritize between a number of promising, but uncertain alternatives. They could be investors deciding between two competing investment portfolios. They could be academics disagreeing over which research method, or theory, will yield the best answers to a research question. They might be collaborating artists, disagreeing over the best way to create an artwork or song. In all cases, Ann and Bob have access to alternative methods of production. Importantly, we maintain the assumption that the players' interests are aligned, to some degree: each of them benefits if the other adopts the best production method --- no matter how strong the disagreement over what such method is.\footnote{For instance, if Ann and Bob are two scientists facing a similar research question, we assume that they share the common goal of pushing knowledge as far as possible in their field.} 

With two technologies, disagreement can be of two kinds. As in section \ref{onetech}, it can be ``vertical,'' when players differ in their degree of optimism.\footnote{This is the case, for instance, if Ann holds \(m^A = (\mathcal{H}_{x},\mathcal{H}_{y})\) while Bob holds \(m^B = (\mathcal{L}_{x},\mathcal{L}_{y})\), or if \(m^A = (\mathcal{H}_{x},\mathcal{H}_{y})\) and \(m^B = (\mathcal{H}_{x},\mathcal{L}_{y})\).} But it can also be ``horizontal,'' when they are optimistic about different technologies --- so that they cannot be ranked in terms of their optimism. For instance, this happens if \(m^A = (\mathcal{H}_{x},\mathcal{L}_{y})\) and \(m^B = (\mathcal{L}_{x},\mathcal{H}_{y})\).

We now show that horizontal disagreement can be a powerful motivator. Proposition \ref{prop4fixed} highlights its benefits when players can be initially assigned to work on a given technology, but are free to pick the desired technology in the second period. In order to state the result formally, let \(\hat{Y}(k^A,k^B,m^A,m^B,Q)\) denote the expected aggregate output of the game when player \(i=A,B\) is constrained to work on technology \(k^i\) in the first period.

\begin{proposition}\label{prop4fixed}
If the alternative technologies are objectively equally productive \((Q_x = Q_y)\), horizontal disagreement will do better than like-mindedness, provided that each player is initially assigned to the technology that she is optimistic about. Formally,
\begin{equation*}
\hat{Y}(x,y,(\mathcal{H}_x,\mathcal{L}_y),(\mathcal{L}_x,\mathcal{H}_y),Q)\ge \hat{Y}(k^A,k^B,m,m,Q)\quad \forall (k^A,k^B,m)\in K^2\times M. 
\end{equation*}
The inequality holds strictly if \(\phi^{m^{-i}}_{m^i}(\cdot,\cdot,x,y|m^i)\) is strictly increasing in \(e^i\).
\end{proposition}
The intuition for the result is as follows. If Ann and Bob start the game like-minded, in the best case scenario they will exert effort \(e^\mathcal{H}\) in every period --- this happens provided that they are optimistic about at least one technology. With initial horizontal disagreement, for any technology assignment, players are motivated to exert more effort in \(t=1\) than statically optimal given the corresponding technology view. By working harder, they increase the probability of their colleague's model change, which, they believe, would ensure that the latter adopts the best technology in \(t=2\). Hence, when initially assigned to the technology they are optimistic about, both Ann and Bob choose effort levels above \(e^\mathcal{H}\). When (\(Q_x = Q_y\)), the increase in effort always translates into higher expected output.

 Why did we need the technology assignment to be exogenous for the result of the previous preposition to hold? The reason is that, when initial models are \(m^A = (\mathcal{H}_x,\mathcal{L}_y)\) and  \(m^B = (\mathcal{L}_x,\mathcal{H}_y)\), it might be easier for Ann to change Bob's mind by using technology \(y\) instead of \(x\), i.e., by proving that Bob's preferred approach does not work. If Ann adopts this strategy, her skepticism towards \(y\) could induce her to pick an effort level below \(e^\mathcal{H}\). To address this concern, we introduce the concept of \textit{equal falsifiability}.

\begin{definition}\label{falsifiability}
Technology views \(\mathcal{H}\) and \(\mathcal{L}\) are equally falsifiable if, for each \(e\in\mathcal{E}\), experiments \((\mathcal{H}(\cdot|e),\mathcal{L}(\cdot|e),\mathcal{Y})\)  and \((\mathcal{L}(\cdot|e),\mathcal{H}(\cdot|e),\mathcal{Y})\) are Blackwell equivalent.
\end{definition}
Equal falsifiability requires that the experiment that draws output from \(\mathcal{H}\) if view \(\mathcal{L}\) is correct and from \(\mathcal{L}\) if \(\mathcal{H}\) is correct is as informative as the original experiment, which draws output from the correct view. In the appendix we provide examples of equally falsifiable views.

\begin{proposition}\label{prop5end} Let the alternative technologies views be equally falsifiable. Then,
\begin{enumerate}[(i)]
    \item When the technologies are equally productive in reality (\(Q_x = Q_y\)), horizontal disagreement will do better than like-mindedness. Formally, 
\begin{equation*}
\hat{Y}((\mathcal{H}_x,\mathcal{L}_y),(\mathcal{L}_x,\mathcal{H}_y), Q)\ge \hat{Y}(m,m,Q)\quad \forall m\in M. 
\end{equation*} 
The inequality holds strictly if \(\phi^{m^{-i}}_{m^i}(\cdot,\cdot,x,y|m^i)\) is strictly increasing in \(e^i\).
\item If the opposing models \((\mathcal{H}_x,\mathcal{L}_y)\) and \((\mathcal{L}_x,\mathcal{H}_y)\)  are equally likely to be correct, a team with horizontal disagreement performs on average better than any like-minded team. Formally, if \(Q = (\mathcal{H}_x,\mathcal{L}_y)\) with probability \(p(\mathcal{H}_x,\mathcal{L}_y) = \frac{1}{2}\) and \(Q = (\mathcal{L}_x,\mathcal{H}_y)\) with probability \(p(\mathcal{L}_x,\mathcal{H}_y) = \frac{1}{2}\), then for each equilibrium profile \(s\in \hat{S}\) it holds
\begin{equation*}
\mathbb{E}_p\left[Y_s((\mathcal{H}_x,\mathcal{L}_y),(\mathcal{L}_x,\mathcal{H}_y), Q)\right]\ge \mathbb{E}_p\left[Y_s(m,m,Q)\right]\quad \forall m\in M. 
\end{equation*}
The inequality holds strictly if \(\phi^{m^{-i}}_{m^i}(\cdot,\cdot,x,y|m^i)\) is strictly increasing in \(e^i\).
\end{enumerate}
\end{proposition}
Proposition \ref{prop5end} (i) states that the intuition discussed in the previous paragraphs holds true with fully endogenous technology choices, if views are equally falsifiable.\footnote{Equal falsifiability is a sufficient condition for the results to hold for any \(\alpha\in[0,1]\).} Disagreement on which technology works best motivates each agent to obtain good outcomes in the first period, in order to convince the co-worker to switch technology. If the technologies work equally well \((Q_x = Q_y\)), the increase in effort always results in higher expected output. The relevance of proposition \ref{prop5end}(ii) is illustrated by the following application.

\begin{paragraph}{Competition of Ideas}
\textit{A manager wants to form a team of two engineers to develop a new product. The product can be based on two very different innovative technologies. Some of the engineers of the firm are optimistic about both alternatives (i.e., hold view \(\mathcal{H}\) for both technologies), others support only one. The manager knows that typically one approach proves better than the other --- one technology is of type \(\mathcal{H}\) and the other of type \(\mathcal{L}\). --- but, from her point of view, both technologies are equally promising ex-ante. Proposition \ref{prop5end} (ii) tells us that the manager will maximize the expected output of the team if she picks co-workers with opposite views, \((\mathcal{H},\mathcal{L})\) vs \((\mathcal{L},\mathcal{H})\), provided that \(\mathcal{H}\) and \(\mathcal{L}\) are equally falsifiable. The benefit of disagreement is twofold: first, both team members will try harder to make more progress during the first production period. Such progress is valuable on its own. Second, the higher first-period effort conveys additional information about the promise of the two technologies, making it more likely that the team will adopt the best technology in the second period.} 
\end{paragraph}
\vspace{.60cm}

The takeaway of this section is simple and yet, we believe, important and non-trivial. When two similarly good production technologies are available, disagreement on the best way to produce might increase group production. We focused on a mechanism that builds on the idea of different perspectives and shows that these differences can be useful even when they do not complement each other, that is, when there is \textit{competition of ideas}. When they have the common goal of increasing each others' output, disagreeing people will challenge each other, and work harder to prove, with their successes, that their perspective is valid, beneficial, and worth adopting. The resulting increase in group output is where --- we believe --- lies the ``disagreement dividend.''

\section{Informativeness, Externalities, and Welfare}\label{discussion}
 The results presented in the previous sections relied on two core forces. First, the presence of production externalities, the driving force of the incentive to facilitate or prevent others from changing their model of production. Second, the idea that, from the point of view of the group of innovators, some actions could be more informative about the true model than others. We have shown that the combination of positive externalities and the assumption that more productive actions are more informative about the underlying production technology can, in many intuitive circumstances, drive up the group's output. 
 
 It seems reasonable to ask ourselves what would happen if we relaxed, or reversed the two assumptions. In the next example, we describe two technology views, \(\mathcal{H}\) and \(\mathcal{L}\), that satisfy all our assumptions except for assumption \ref{ass3}: these views are, in fact, such that more effort conveys less information about which one is correct.

\begin{example}\label{ex4}
View \(\mathcal{H}\) is such that \(Y^i = \gamma_0 + \gamma_1e^i + \varepsilon\) where \(e^i\in [0,b]\), \(\varepsilon \sim N(0,\sigma^2)\) independent of \(e^i\), and \(\gamma_0,\gamma_1>0\). View \(\mathcal{L}\) is such that \(Y = \gamma_2e^i + u\) where \(e^i\in [0,b]\), \(\varepsilon \sim N(0,\sigma^2)\) independent of \(e^i\), and \(\gamma_2>\gamma_1\) such that \(\gamma_2b<\gamma_0 + \gamma_1b\).
\end{example}
To gain an intuition for why low effort is more informative than high effort in the case of example \ref{ex4}, note that the restrictions on \(\gamma_0,\gamma_1\) and \(\gamma_2\) imply that the predictions of the two views differ the most when no effort is invested, that is, \(e = 0\), and converge as effort increase (while noise does not depend on effort). If we instead had \(\gamma_1>\gamma_2>0\), the two technology views would satisfy assumption \ref{ass3}.

Relaxing assumption \ref{ass3}, as in our previous example, 
changes players' equilibrium behavior dramatically. The probability \(\phi^{m^{-i}}_{m^i}(\cdot,\mathbf{k}|m^i)\) becomes decreasing in effort. As a consequence, an agent who wants to persuade her disagreeing colleague to change views will exert less effort in the first period; one who wants disagreement to persist will instead work harder. In the two-technology case, the implication is that horizontal disagreement will push each player to work less than if the other player was like-minded. When the two technologies are ex-ante equivalent in reality, no team will produce more, in expectation, than like-minded players agreeing on \((\mathcal{H}_x,\mathcal{H}_y)\).  

A similarly crucial role is played by the assumption that each team member is \textit{ceteris paribus} better off when other members produce more. The positive externality creates the incentive to persuade disagreeing co-workers to abandon skeptical views, as well as the one to convince them to adopt the best technology. Clearly, if other players' output entered utility negatively, the incentive would be very different. To see this, consider equations \ref{exintro}, \ref{exintro2} and \ref{exintro3} of our illustrative example of section \ref{illustration}, but let \(\beta<0\) --- so that externalities are negative. By inspecting the equations, it is easy to see that \(\beta<0\) implies that the value of triggering a co-worker's change in mind at the end of the first period becomes negative: as a result, disagreement reduces Ann's first-period effort and expected output, rather than increasing them.

While reversing assumption \ref{ass3} or the positive externality assumption alone changes our results drastically, reversing both forces simultaneously leaves our main results unchanged.

\begin{claim}\label{claim1}
The main results of this paper remain valid if the following changes are made to the primitives and assumptions of the model 
\begin{enumerate} [(i)]
    \item Externalities from production are negative, \(\frac{\partial v(y^{-i})}{\partial y^{{-i}}}<0\) 
    \item For each \(e^{\prime},e\in \mathcal{E}\), action \(e^\prime\) is more informative than \(e\) if and only if \(e^\prime<e\).
\end{enumerate}
\end{claim}
We omit the formal proofs for the claim, as such proofs would follow very closely the ones reported in the appendix. Instead, we provide the intuition for the one-technology case (the multiple-technology case follows the same intuition). Let the alternative assumptions of claim \ref{claim1} hold, and consider a disagreeing team. Note that the player with view \(\mathcal{H}\) will be hurt from a change of mind of the \(\mathcal{L}\)-view player: by assumption \ref{ass2}, the skeptic works harder after becoming optimist, damaging her co-worker through the negative production externalities of point (i). This force pushes the optimistic player to reduce information arrival in the first period, in order to decrease the (subjective) probability of a co-worker's change in mind. However, when the relation between effort and informativeness is inverse --- as of point (ii) --- the way to reduce information arrival is by working harder.  Hence, even under these alternative assumptions, disagreement pushes optimists to work harder and, if negative externalities are strong enough, a disagreeing team will produce on average more than any like-minded team.\footnote{A few details of the propositions will indeed need to be intuitively modified for the result to hold under the assumptions of claim \ref{claim1}. In particular, the second part of proposition \ref{prop3} (ii) holds for \(\Delta\) \textit{negative} enough; and the requirement for inequalities to be strict becomes that \(\phi^{m^{-i}}_{m^i}(\mathbf{e},\mathbf{k}|m^i)\) is strictly \textit{decreasing} in effort.} Regardless of this equivalence, we believe that our original specification --- with positive externalities and more information arriving the more a team works on a project --- is particularly realistic and captures well the zest of most of our motivating examples. 

We conclude the discussion with a word on the welfare implications of disagreement. A welfare analysis would be particularly complex in our setup, which does not impose strong assumptions on the true process \(Q\). Without such assumptions, it is hard to tell, for instance, whether the effort levels that maximize team members' joint expected payoffs (or are Pareto efficient in terms of expected utility) are above or below \(e^{\mathcal{H}}\) and \(e^{\mathcal{L}}\). What we can say with certainty is that if the assumptions of proposition \ref{prop5end} part (i) hold and if, additionally, \(Q_x=Q_y=\mathcal{H}\),  any Pareto efficient stage effort must be above \(e^\mathcal{H}\). Hence the boost in players' effort generated by horizontal disagreement can --- as in the example presented in section \ref{illustration} --- lead to a Pareto improvement. Not much can be concluded, however, in general.\footnote{For instance, note that with \(|K|=1\) and \(Q=\mathcal{H}\), the expected payoff of an optimist is always higher when she's paired with another optimist. In such case, disagreement cannot lead to a Pareto improvement in terms of expected payoffs.}

This final observation leaves us with a word of caution: we have shown that disagreement can increase effort. We can expect this to boost innovation and output. Increasing innovation might be the goal of a team manager, or of society, especially if the breakthroughs and innovations will prove largely beneficial for many and these benefits are not internalized by the team. From the point of view of team members, however, the cost of disagreement could be very high.

\section{Conclusion} \label{exit}

We have shown that model disagreement within a group of economic agents who repeatedly engage in a productive activity can increase the aggregate output of the group. We have unveiled a relation between externalities, disagreement and productivity, which helps us benchmark our results with the main findings of the theoretical literature about diversity in teams of problem solvers \citep[e.g.,][]{Hong2001,Hong2004}. In this literature, different perspectives are seen as an asset, but under two assumptions. First, team members must be able to cooperate and combine their perspectives in a productive way. Second, and related, different perspectives must not lead to different goals. 

In the paper, we have presented a mechanism that in surface departs from the first assumption: even when different perspectives (models) are in\textit{ conflict} with each other, leading to disagreement about the most productive approach, they might still be very useful to a team of innovators. Peers' skepticism and competition of ideas can motivate team members to work harder to prove their point. If anything, our results illustrate that some degree of ``scientific'' skepticism of each group member towards the perspective of others can be a powerful motivator: the disagreeing group should be aware that only perspectives that prove successful are eventually adopted, as this force might push them to work harder to convince others. At the same time, from a high-level point of view, our findings suggest that the benefit of disagreement should materialize if agents' ultimate goals are somewhat aligned by positive production externalities, while differences might harm if players want to reduce the productivity of others (negative externalities). This is take-away is in line with the idea that a ``diversity premium'' relies on diverse people ultimately working towards similar goals.

Finally, from a theoretical standpoint, we have discussed a type of communication that --- we believe --- is somewhat overlooked by the economic literature: the sort of persuasion that comes from the tangible results of economic actions, rather than from information design by a sender or the implicit informational content of a given equilibrium behavior. We believe that a deeper investigation of this force could help explain a variety of phenomena that go beyond productive incentives, including voter polarization, excessive debt accumulation, and other policy distortions.  

\bibliographystyle{apalike}
\bibliography{biblio}

\begin{thebibliography}{}

\bibitem[Alesina and La~Ferrara, 2005]{AlesinaLaF2005}
Alesina, A. and La~Ferrara, E. (2005).
\newblock Ethnic diversity and economic performance.
\newblock {\em Journal of Economic Literature}, 43(3):762--800.

\bibitem[Alesina and Tabellini, 1990]{AlesinaT1990}
Alesina, A. and Tabellini, G. (1990).
\newblock A positive theory of fiscal deficits and government debt.
\newblock {\em The Review of Economic Studies}, 57(3):403--414.

\bibitem[Andreoni and Mylovanov, 2012]{Andreoni2012}
Andreoni, J. and Mylovanov, T. (2012).
\newblock Diverging opinions.
\newblock {\em American Economic Journal: Microeconomics}, 4(1):209--32.

\bibitem[Arrow, 1951]{Arrow1951}
Arrow, K.~J. (1951).
\newblock {\em Social Choice and Individual Values}.
\newblock (Cowles Commission Monogr. No. 12) Wiley, Chicago.

\bibitem[Ba, 2022]{Ba2022}
Ba, C. (2022).
\newblock {Robust Model Misspeciﬁcation and Paradigm Shifts}.
\newblock {\em SSRN Electronic Journal}.

\bibitem[Blackwell and Girshick, 1962]{Blackwell1954}
Blackwell, D. and Girshick, M.~A. (1962).
\newblock {\em Theory of Games and Statistical Decisions}.
\newblock Wiley.

\bibitem[Center and Bates, 2009]{TechPolitik}
Center, S. and Bates, E. (2009).
\newblock Tech-politik: Historical perspectives on innovation, technology and strategic competition.
\newblock {\em CSIS Briefings}.

\bibitem[Che and Kartik, 2009]{Che2009}
Che, Y. and Kartik, N. (2009).
\newblock Opinions as incentives.
\newblock {\em Journal of Political Economy}, 117(5):815--860.

\bibitem[Crawford and Sobel, 1982]{CS}
Crawford, V.~P. and Sobel, J. (1982).
\newblock Strategic information transmission.
\newblock {\em Econometrica}, 50(6):1431--1451.

\bibitem[Dong, 2018]{Dong2018}
Dong, M. (2018).
\newblock Strategic experimentation with asymmetric information.
\newblock {\em Mimeo}.

\bibitem[Dong and Mayskaya, 2022]{Dong2022}
Dong, M. and Mayskaya, T. (2022).
\newblock Does reducing communication barriers promote diversity?
\newblock {\em Working Paper}.

\bibitem[Esponda and Pouzo, 2016]{Esponda2016}
Esponda, I. and Pouzo, D. (2016).
\newblock {Berk–Nash Equilibrium: A Framework for Modeling Agents With Misspecified Models}.
\newblock {\em Econometrica}, 84(3):1093--1130.

\bibitem[Evans, 1990]{Evans1990}
Evans, J. S. B.~T. (1990).
\newblock {\em Bias in Human Reasoning: Causes and Consequences}.
\newblock Psychology Press.

\bibitem[Fudenberg and Lanzani, 2022]{Fudenberg}
Fudenberg, D. and Lanzani, G. (2022).
\newblock {Which misspecifications persist?}
\newblock {\em Theoretical Economics}.

\bibitem[Galperti, 2019]{Galperti2019}
Galperti, S. (2019).
\newblock {Persuasion: The Art of Changing Worldviews}.
\newblock {\em American Economic Review}, 109(3):996--1031.

\bibitem[Grant, 2021]{Grant2021}
Grant, A. (2021).
\newblock Persuading the unpersuadable.
\newblock {\em Harvard Business Review}.

\bibitem[Hart and Rinott, 2020]{Hart2020}
Hart, S. and Rinott, Y. (2020).
\newblock {Posterior probabilities: Dominance and optimism}.
\newblock {\em Economics Letters}, 194:109352.

\bibitem[Heidhues et~al., 2018]{Heidues2018}
Heidhues, P., Kőszegi, B., and Strack, P. (2018).
\newblock Unrealistic expectations and misguided learning.
\newblock {\em Econometrica}, 86(4):1159--1214.

\bibitem[Hirsch, 2016]{Hirsch2016}
Hirsch, A.~V. (2016).
\newblock Experimentation and persuasion in political organizations.
\newblock {\em American Political Science Review}, 110(1):68–84.

\bibitem[Hong et~al., 2007]{Hong2007}
Hong, H., Stein, J.~C., and Yu, J. (2007).
\newblock {Simple Forecasts and Paradigm Shifts}.
\newblock {\em Journal of Finance}, LXII(3).

\bibitem[Hong and Page, 2001]{Hong2001}
Hong, L. and Page, S.~E. (2001).
\newblock {Problem Solving by Heterogeneous Agents}.
\newblock {\em Journal of Economic Theory}, 97(1):123--163.

\bibitem[Hong and Page, 2004]{Hong2004}
Hong, L. and Page, S.~E. (2004).
\newblock Groups of diverse problem solvers can outperform groups of high-ability problem solvers.
\newblock {\em Proceedings of the National Academy of Sciences}, 101(46):16385--16389.

\bibitem[Janis, 1982]{Janis82}
Janis, I.~L. (1982).
\newblock {\em Groupthink: psychological studies of policy decisions and fiascoes}.
\newblock Houghton Mifflin, Boston.

\bibitem[Kartik et~al., 2021]{Kartik2021}
Kartik, N., Lee, F.~X., and Suen, W. (2021).
\newblock {Information Validates the Prior: A Theorem on Bayesian Updating and Applications}.
\newblock {\em American Economic Review: Insights}, 3(2):165--82.

\bibitem[Keller et~al., 2005]{Keller2005}
Keller, G., Rady, S., and Cripps, M. (2005).
\newblock {Strategic Experimentation with Exponential Bandits}.
\newblock {\em Econometrica}, 73(1):39--68.

\bibitem[Kuhn, 1962]{Kuhn1962}
Kuhn, T.~S. (1962).
\newblock {\em The Structure of Scientific Revolutions}.
\newblock University of Chicago Press, Chicago.

\bibitem[Lewis et~al., 2019]{Lewis}
Lewis, J., Fraga, K., and Erickson, T. (2019).
\newblock {\em Dichotomous Thinking. In: Zeigler-Hill, V., Shackelford, T. (eds) Encyclopedia of Personality and Individual Differences}.
\newblock Springer, Cham.

\bibitem[Merchant, 2018]{iPhone}
Merchant, B. (2018).
\newblock {\em The One Device: The Secret History of the IPhone}.

\bibitem[Neyman et~al., 1933]{NeymanPearson}
Neyman, J., Pearson, E.~S., and Pearson, K. (1933).
\newblock Ix. on the problem of the most efficient tests of statistical hypotheses.
\newblock {\em Philosophical Transactions of the Royal Society of London. Series A, Containing Papers of a Mathematical or Physical Character}, 231(694-706):289--337.

\bibitem[Nickerson, 1998]{Nickerson1998}
Nickerson, R. (1998).
\newblock Confirmation bias: A ubiquitous phenomenon in many guises.
\newblock {\em Review of General Psychology}, 2:175--220.

\bibitem[Ortoleva, 2012]{Ortoleva2012}
Ortoleva, P. (2012).
\newblock Modeling the change of paradigm: Non-bayesian reactions to unexpected news.
\newblock {\em American Economic Review}, 102(6):2410--2436.

\bibitem[Page, 2007]{Page}
Page, S.~E. (2007).
\newblock {\em The Difference: How the Power of Diversity Creates Better Groups, Firms, Schools, and Societies (New Edition)}.
\newblock Princeton University Press.

\bibitem[Prat, 2002]{Prat2002}
Prat, A. (2002).
\newblock {Should a team be homogeneous?}
\newblock {\em European Economic Review}, 46(7):1187--1207.

\bibitem[Rees-Jones and Taubinsky, 2019]{Rees-Jones}
Rees-Jones, A. and Taubinsky, D. (2019).
\newblock {Measuring “Schmeduling”}.
\newblock {\em The Review of Economic Studies}, 87(5):2399--2438.

\bibitem[Salmela and Oikkonen, 2022]{Metallica}
Salmela, E. and Oikkonen, E. (2022).
\newblock How conflicts with partners help an underdog to innovate and grow the largest in the world-case metallica.
\newblock {\em Conflict Resolution Quarterly}, 40(1):141--158.

\bibitem[Schwartzstein and Sunderam, 2021]{Schwartzstein2021}
Schwartzstein, J. and Sunderam, A. (2021).
\newblock {Using Models to Persuade}.
\newblock {\em American Economic Review}, 111(1):276--323.

\bibitem[Scott, 2017]{Scott2017}
Scott, K. (2017).
\newblock What steve jobs taught me about debate in the workplace.
\newblock {\em NBC News}.

\bibitem[Shapiro and Varian, 1999]{ShapiroVarian}
Shapiro, C. and Varian, H.~R. (1999).
\newblock The art of standards wars.
\newblock {\em California Management Review}, 41(2):8--32.

\bibitem[Torgersen, 1970]{Torgersen}
Torgersen, E. (1970).
\newblock Comparison of experiments when the parameter space is finite.
\newblock {\em Z. Wahrscheinlichkeitstheorie verw Gebiete}, 16:219–249.

\bibitem[Van~den Steen, 2010]{VanDerSteen}
Van~den Steen, E. (2010).
\newblock Culture clash: The costs and benefits of homogeneity.
\newblock {\em Management Science}, 56(10):1718--1738.

\end{thebibliography}
\newpage
\appendix
\section{Appendix}
\vspace{.5cm}

\subsection{Examples}

\begin{example}[Discrete Bandit]\label{ex1} According to view \(\mathcal{L}\), \(Y^i = r\ge0\) with probability \(F(e^i)\) and \(Y^i = 0\) with probability \(1-F(e^i)\), where \(F:\mathcal{E}\to[0,1]\) is differentiable and strictly increasing. According to view \(\mathcal{H}\), \(Y^i = R>r\) with probability \(F(e^i)\) and \(Y^i = 0\) with probability \(1-F(e^i)\). For both models, \(F(0)=0\).
\end{example}
When \(r=0\), the structure of returns recalls exponential bandit problems \citep{Keller2005}, where view \(\mathcal{L}\) is equivalent to the technology being a ``bad arm,'' and \(\mathcal{H}\) describing a ``good arm.''

\begin{example}[Log-concave Noise]\label{ex2} Each technology view \(m_k\in{\mathcal{H},\mathcal{L}}\) takes the form \(Y^i = \varphi(e^i,m_k) + \varepsilon^i_k\). The function \(\varphi: \mathcal{E}\times \{\mathcal{H},\mathcal{L}\}\to \mathbb{R}\) is differentiable and increasing in \(e^i\), \(\varphi(e^i,\mathcal{H})-\varphi(e^i,\mathcal{L})\) is strictly increasing in \(e^i\), \(\varphi(0,m_k) = 0\), and \(\varepsilon^i_k\) is white noise with a log-concave continuous probability distribution, independent of \(e^i\) and \(m_k\). 
\end{example}
The characteristics of the views of example \ref{ex2} --- in particular, the supermodularity of expected output and noise log-concavity --- recall the returns structure of \citet{Heidues2018}.

\begin{remark} The views of example \ref{ex2} are equally falsifiable if the distribution of \(\varepsilon\) is symmetric. The views of example \ref{ex1} are equally falsifiable if \(r>0\).
\end{remark}

\begin{example}\label{ex3} Let the true production process \(Q\) be such that \(\mathbb{E}_Q[Y^i|e^i]=\gamma e^i\), \(\gamma>0\). Production model \(\mathcal{H}\) is \(Y^i = \gamma_{\mathcal{H}} e^i + \varepsilon\), where \(\varepsilon \sim U[-\psi, \psi], \gamma_{\mathcal{H}}>0, \psi>0\). Model \(\mathcal{L}\) specification is \(Y^i = u\), \(u \sim U[-\psi, \psi]\). \(\psi\) is large relative to \(\gamma_{\mathcal{H}}\). Stage utility is \(U^i = y^i + \beta y^{-i} - \frac{1}{2}(e^i)^2\), where \(\beta>0\). Note that if both Ann and Bob start the game with model \(\mathcal{H}\), each of them exerts effort \(e^i = \gamma_{\mathcal{H}}\) in every period, so that \(\hat{Y}(\mathcal{H},\mathcal{H}, Q) = 4\gamma\gamma_{\mathcal{H}}\). Consider now the game where Ann starts with model \(\mathcal{H}\) and Bob starts with model \(\mathcal{L}\). From Ann's perspective, Bob's switch from \(\mathcal{L}\) to \(\mathcal{H}\) at the end of \(t=1\) is worth \(\beta\gamma_{\mathcal{H}}^2\), in expectation. By applying rule \ref{LR} with the null hypothesis that \(Q = m^A_1\), Bob will switch model from \(\mathcal{L}\) to \(\mathcal{H}\) with probability \(\alpha\) if \(y^A_1\in[\gamma_{\mathcal{H}}e_1^A-\psi,\psi]\), with unit probability if \(y^A_1\in[\psi,\gamma_{\mathcal{H}}e_1^A+\psi]\), and will not switch otherwise. It is easy to see that Ann reckons that the probability that Bob will switch is \(\alpha + \frac{\gamma_\mathcal{H}}{2\psi}e^A_1\) for \(e^A_1\in[0,\frac{2\psi}{\gamma_{\mathcal{H}}}(1-\alpha)]\), reaches \(1\) at \(e^A_1 = \frac{2\psi}{\gamma_{\mathcal{H}}}(1-\alpha)\), and remains \(1\) at higher levels of effort. Hence, in period \(t=1\), she chooses \(e^A_1\) to maximize \(\gamma_{\mathcal{H}} e^A_1 + \left[\mathbbm{1}(e^A_1\le\frac{2\psi}{\gamma_{\mathcal{H}}}(1-\alpha))\frac{\gamma_{\mathcal{H}} e^A_1}{2\psi}+ \mathbbm{1}(e^A_1>\frac{2\psi}{\gamma_{\mathcal{H}}}(1-\alpha))\right]\beta\gamma_{\mathcal{H}}^2-\frac{1}{2}(e^A_1)^2\). Note that if \(\beta\to 0\), equilibrium \(e^A_1\) tends to \(\gamma_{\mathcal{H}}\), so that \(\hat{Y}(\mathcal{H},\mathcal{L},Q)<\hat{Y}(\mathcal{H},\mathcal{H},Q)\). On the other hand, if \(\beta \to \infty\), equilibrium effort tends to \(\frac{2\psi}{\gamma_{\mathcal{H}}}(1-\alpha)\). As a consequence, if \(\frac{2\psi}{\gamma_{\mathcal{H}}}(1-\alpha)>4\gamma_{\mathcal{H}}\), it holds that \(\hat{Y}(\mathcal{H},\mathcal{L},Q)>\hat{Y}(\mathcal{H},\mathcal{H},Q)\) for \(\beta\) large enough.
\end{example}

\subsubsection{Effort and Informativeness in Example \ref{ex1} and Example \ref{ex2}}

We show that the technology views of examples \ref{ex1} and \ref{ex2} satisfy assumption \ref{ass3}.
Fix a technology \(k\), and let \(\mathcal{H}\) and \(\mathcal{L}\) be the technology views of example \ref{ex1}. Let \(\mathcal{Y}= \{R,r,0\}\). We want to show that for all \(e, e^\prime\in \mathcal{E}\), experiment \(\Pi_{e^\prime}\) is more informative than \(\Pi_{e}\) if \(e^\prime>e\). Consider \(e, e^\prime\in \mathcal{E}\), and let \(e^\prime>e\). \(\Pi_{e^{\prime}}\) is such that if \(Q_k = \mathcal{H}\) then \(Y = R\) with probability \(F(e^\prime)\) and \(Y = 0\) with remaining probability; if \(Q_k = \mathcal{L}\) then \(Y = r\) with probability \(F(e^\prime)\) and \(Y = 0\) with remaining probability.  Similarly, \(\Pi_e\) is such that if \(Q_k = \mathcal{H}\) then \(Y = R\) with probability \(F(e)\) and \(Y = 0\) with remaining probability; if \(Q_k = \mathcal{L}\) then \(Y = r\) with probability \(F(e)\) and \(Y = 0\) with remaining probability. \(\Pi_e\) can be obtained from \(\Pi_{e^\prime}\) by applying the garbling \(g: \mathcal{Y}\to \Delta(\mathcal{Y})\), where \(g(0|0) = 1\), \(g(R|R) = g(r|r) = \frac{F(e)}{F(e^\prime)}\) and \(g(0|R) = g(0|r) = 1- \frac{F(e)}{F(e^\prime)}\). Hence, \(\Pi_{e^\prime}\) is strictly more informative than \(\Pi_{e}\).

Consider now the technology view of example \ref{ex2}. To show that \(\Pi_{e^\prime}\) is (strictly) more informative than \(\Pi_{e}\) we exploit the following characterization of Blackwell informativeness \citep{Blackwell1954}. 

\begin{lemma}[Power and Informativeness]\label{blackwell}
Fix a binary state space \(\Omega = \{\omega_0,\omega_1\}\) and a true state \(\hat{\omega}\in\Omega\). Let \(\omega, \omega^\prime\in \Omega\), \(\omega\ne \omega^\prime\) and consider testing the null hypothesis \(\hat{\omega} = \omega\) against \(\hat{\omega} = \omega^\prime\).  Experiment \(G\) is more informative than experiment \(P\) if and only if, for all \(\alpha\in [0,1]\), the most powerful size-\(\alpha\) test based on experiment \(G\) is at least as powerful as the most powerful size-\(\alpha\) test based on experiment \(P\).  
\end{lemma}
The result is proven in \citet{Blackwell1954}, chapter 12, and by \citet{Torgersen} in a more general version. To see that in example \ref{ex2} \(\Pi_{e^\prime}\) is (strictly) more informative than \(\Pi_{e}\) it suffices to show that, for any \(\alpha\in [0,1]\), the most powerful test of any technology view against the other has weakly larger power when the data are obtained from \(\Pi_{e^\prime}\) than when they are obtained from \(\Pi_{e}\). Let the null hypothesis be \(Q_k = \mathcal{L}\) and the alternative be \(Q_k = \mathcal{H}\). According to the null \(Y = \varphi(e,\mathcal{L}) + \varepsilon\), where \(\varepsilon\) is independent of \(e\) and the true model, and has the monotone likelihood ratio property (due to the log-concavity of its distribution \(F_{\varepsilon}\)). We show the result for cases where the likelihood ratio is strictly monotone on \(\mathbb{R}\); the proof for the case of weak monotonicity follows the same logic but is more tedious. Fix \(\alpha\in [0,1]\). If \(\alpha\in\{0,1\}\), then the power of the test is either \(0\) or \(1\), independent of effort. If \(\alpha\in (0,1)\), by \citet{NeymanPearson}, the most powerful test for the null hypothesis given \(e\) and \(y\) rejects \(Q_k = \mathcal{L}\) if and only if 
\begin{equation*}
    y>y_\alpha
\end{equation*}
for \(y_\alpha = F^{-1}_{\varepsilon}(1-\alpha) + \varphi(e,\mathcal{L})\), so that the power of the test is \(\phi^\mathcal{L}_{\mathcal{H}}(e|\mathcal{H}) = 1-F_{\varepsilon}(F^{-1}_{\varepsilon}(1-\alpha) + \varphi(e,\mathcal{L})-\varphi(e,\mathcal{H}))\). Now, note that \(F_{\varepsilon}\) is increasing, while \(\varphi(e,\mathcal{L})-\varphi(e,\mathcal{H})\) is decreasing in \(e\) by assumption on \(\varphi\), hence \(\phi^\mathcal{L}_{\mathcal{H}}(\cdot|\mathcal{H})\) is increasing. This implies that if \(e^\prime>e\) the most powerful test based on \(\Pi_{e^\prime}\) is more powerful than the most powerful test based on \(\Pi_e\). Following similar steps, one can show that power is increasing on \(\mathcal{E}\) also when the null hypothesis is \(Q_k = \mathcal{H}\). Hence \(\Pi_{e^\prime}\) is more informative than \(\Pi_{e}\) by lemma \ref{blackwell}.

\subsection{Proof of Lemma \ref{lemmapower}} Let \(|K| = 1\). Let \(e^A,e^B\in \mathcal{E}\), and consider the experiment \(\Pi_{(e^A,e^B)} = (\Pi_{e^A},\Pi_{e^B})\) obtained if Ann and Bob choose \(e^A\) and \(e^B\), respectively, in period \(t=1\). Consider any \(e^\prime\in \mathcal{E},e^\prime>e^A\). By assumption \ref{ass3}, \(\Pi_{e^\prime}\) is more informative than \(\Pi_{e^A}\). Hence, by known results in \citep{Blackwell1954}, there must exist a garbling \(g: \mathcal{Y}\to \Delta(\mathcal{Y})\) such that \(\Pi_{e^A}\) is replicated applying \(g\) to the signal realizations produced by \(\Pi_{e^\prime}\). It holds that the joint likelihood of any signal realization from \(\Pi_{(e^A,e^B)}\) if \(\mathcal{H}\) is the true view is
\begin{equation*}
L_{\mathcal{H}}(y^A,y^B|e^A, e^B) = \mathcal{H}(y^B|e^B)\sum_{y\in\mathcal{Y}}g(y^A|y)\mathcal{H}(y|e^\prime)
\end{equation*}
where we used the assumption of the independence of output realizations across players. Similarly, if \(\mathcal{L}\) is the true view, we have 
\begin{equation*}
L_{\mathcal{L}}(y^A,y^B|e^A, e^B) = \mathcal{L}(y^B|e^B)\sum_{y\in\mathcal{Y}}g(y^A|y)\mathcal{L}(y|e^\prime).\footnote{Clearly summation sign can be replaced by integration in the case of continuous random variables.}
\end{equation*}
We have shown that we can induce experiment \(\Pi_{(e^A,e^B)}\) by taking \(\Pi_{(e^\prime,e^B)}\) and applying garbling \(g\) to signal realizations \(y^\prime\) from \(\Pi_{e^\prime}\). Hence \(\Pi_{(e^\prime,e^B)}\) is more informative than \(\Pi_{(e^A,e^B)}\). With the same procedure, one can show that, if \(e^\prime>e^B\), \(\Pi_{(e^A, e^\prime)}\) is more informative than \(\Pi_{(e^A,e^B)}\). Hence the informativeness of \(\Pi_{(e^A,e^B)}\) is increasing in \(e^i\), \(i=A,B\).

Now, fix \(\alpha\in [0,1]\), any \(m^i,m^{-i}\in M\), and any \(e^A,e^B\in \mathcal{E}\). Consider the size-\(\alpha\) likelihood ratio test of the null hypothesis \(Q = m^{-i}\) against \(Q = m^{i}\), when effort choices are \(e^A,e^B\in \mathcal{E}\). For any given experiment, it follows from \citep{NeymanPearson} that the likelihood ratio test is the (uniformly) most powerful among exact tests for dichotomies. Hence \(\phi^{m^{-i}}_{m^i}(e^A,e^B|m^i)\) is the maximum power attainable by the test based on  \(\Pi_{(e^A,e^B)}\). By lemma \ref{blackwell}, for any \(e^\prime,e\in \mathcal{E}\), 
\begin{equation*}
    \Pi_{(e,e^\prime)}\ge \Pi_{(e^A,e^B)} \implies \phi^{m^{-i}}_{m^i}(e,e^\prime|m^i)\ge \phi^{m^{-i}}_{m^i}(e^A,e^B|m^i)
\end{equation*}
where \(\Pi_{(e,e^\prime)}\ge \Pi_{(e^A,e^B)}\) indicates that  \(\Pi_{(e,e^\prime)}\) is more informative than \(\Pi_{(e^A,e^B)}\). But we have already shown that \(e^\prime>e^A\implies \Pi_{(e^\prime, e^B)}\ge \Pi_{(e^A,e^B)}\) and \(e^\prime>e^B\implies \Pi_{(e^A, e^\prime)}\ge \Pi_{(e^A, e^B)}\). Hence \(\phi^{m^{-i}}_{m^i}(e^A,e^B|m^i)\) is increasing in \(e^A\) and \(e^B\), proving lemma \ref{lemmapower}.

\subsection{Benchmark: Unawareness of Disagreement and Myopia}
In this section, we derive a benchmark for proposition \ref{prop2formation}

\begin{definition}\label{aware}
We say that player \(i\in I\) with initial model \(m^i_1\in M\) is unaware of disagreement if she always assumes that the profile of initial models \(\mathbf{m}_1\) is \((m^i_1,m^i_1)\). We say that players are myopic about disagreement if equilibrium condition (ii) is satisfied for \(\delta = 0\). 
\end{definition}
Ann is unaware of disagreement if she always thinks that Bob shares her same model. Unawareness of disagreement at time \(t=1\) means that Ann does not conceive alternative models, so that \(m^A_1 = m^A_2\) with certainty. If myopic, Ann does not account for how her output might affect Bob's future model choice. However, she is aware of Bob's current model and switches to it if persuaded by evidence. Finally, we denote by \(\hat{Y}_u(m^A,m^B,Q)\) the expected aggregate output when Ann and Bob are unaware of disagreement, and \(\hat{Y}_{o}(m^A,m^B,Q)\) the one when they are myopic, assuming that they start the game with models \(m^A,m^B\in M\). In both cases, players play according to equilibrium strategies, provided that their discount rate and perception of the state are modified as in definition \ref{aware}.

\begin{proposition}[Benchmark]\label{prop1bench}
The following results hold:
\begin{enumerate}[(i)]
    \item For each \(m\in M\), \(\iota\in \{u,o\}\), it holds
      \begin{equation*}
    \hat{Y}_{\iota}(m,m,Q) = \hat{Y}(m,m,Q).
    \end{equation*}
    \item If Ann and Bob are unaware of disagreement, it holds
    \begin{equation*}
    2\hat{Y}_u(\mathcal{H},\mathcal{L},Q) = \hat{Y}_{u}(\mathcal{H},\mathcal{H},Q) + \hat{Y}_{u}(\mathcal{L},\mathcal{L},Q).
    \end{equation*}
    If instead Ann and Bob are myopic, it holds,
    \begin{align*}
    2\hat{Y}_o(\mathcal{H},\mathcal{L},\mathcal{H}) &\ge \hat{Y}_o(\mathcal{H},\mathcal{H},\mathcal{H}) + \hat{Y}_o(\mathcal{L},\mathcal{L},  \mathcal{H})\\
    2\hat{Y}_o(\mathcal{H},\mathcal{L},\mathcal{L}) &\le \hat{Y}_o(\mathcal{H},\mathcal{H},\mathcal{L}) + \hat{Y}_o(\mathcal{L},\mathcal{L},\mathcal{L}),
    \end{align*}
    The first inequality holds strictly if \(\phi_{m^{i}}^{m^{-i}}(\cdot,\cdot|m^i)\) is strictly increasing in \(e^A\) and \(e^B\).
    \item For \(\iota\in\{u,o\}\), \(m,m^\prime\in M\),
    \begin{equation*}
     \hat{Y}_{\iota}(\mathcal{H},\mathcal{H},Q)>\hat{Y}_{\iota}(m,m^\prime,Q) \iff (m,m^\prime)\ne (\mathcal{H},\mathcal{H}).
    \end{equation*}
\end{enumerate}
\end{proposition}

\paragraph{Proof of Proposition \ref{prop1bench}}
Consider part (i). By the separability of the payoff function at \(t=2\) each agent's equilibrium effort only depends on their own model. For \(i=A,B\), and \(m^A_2,m^{B}_2\in M\), \(s^i_{2e}(m^A_2,m^B_2)\) is the solution to the problem
\begin{equation}\label{maxperiod2}
\max_{e\in \mathcal{E}} \mathbb{E}_{m^i_2}[u(Y,e)|e].
\end{equation}
By assumption \ref{ass2}, the problem has a unique solution, equal to \(e^\mathcal{H}\) if \(m^i_2 = \mathcal{H}\) and \(e^\mathcal{L}\) if \(m^i_2 = \mathcal{L}\). Next, for \(m\in\{\mathcal{H},\mathcal{L},Q\}\), define \(y^{\mathcal{H}}_m = \mathbb{E}_m[Y|e^{\mathcal{H}}]\) and \(y^{\mathcal{L}}_m = \mathbb{E}_m[Y|e^{\mathcal{L}}]\). Note that assumptions \ref{ass1} and \ref{ass2} imply \(e^{\mathcal{H}}> e^{\mathcal{L}}\) and
\(y_m^{\mathcal{H}}\ge  y_m^{\mathcal{L}}\) for all \(m\in\{\mathcal{H},\mathcal{L},Q\}\), with \(y_m^{\mathcal{H}}> y_m^{\mathcal{L}}\) if \(m\in\{\mathcal{H},Q\}\), by the strict FOSD-monotonicity assumptions.

Next, fix any \(m^\star\in M\). If initial models are \(m^A_1=m^B_1=m^\star\) then when players test their models at the end of the period, the null hypothesis model \(m^i_1\) and the alternative model \(m^{-i}_1\) coincide, for any \(i= A,B\). Hence both players will hold \(m^\star\) in \(t = 2\) with certainty.

Consequently, at \(t=1\), \(i\)'s maximization problem is 

\begin{align*}
\max_{e\in\mathcal{E}} \mathbb{E}_{m^\star}\left[U^i(y,e)|e,s_1^{-i}(m^\star,m^\star)\right] + \delta^i V^i_{s_2,m^\star}(m^\star,m^\star),
\end{align*}
which admits the same solution as
\begin{align}
\max_{e\in\mathcal{E}} \mathbb{E}_{m^\star}\left[u(y,e)|e\right] \label{maxperiod1}
\end{align}
because \(V^i_{s_2,m^\star}(m^\star,m^\star)\) and \(v\) do not depend on \(e\).
So, for each \(i=A,B\), agreement at \(t=1\) means that in equilibrium 
\begin{gather}
s^i_{1e}(\mathcal{H},\mathcal{H})=s^i_{2e}(\mathcal{H},\mathcal{H}) = e^\mathcal{H} \label{like-minded1} \\
s^i_{1e}(\mathcal{L},\mathcal{L})= s^i_{2e}(\mathcal{L},\mathcal{L}) = e^\mathcal{L}. \label{like-minded2}
\end{gather}
Evaluating aggregate expected output across periods and players, we obtain \(\hat{Y}(m^\star,m^\star, Q) = 4\mathbb{E}_Q[Y|e^{\mathcal{H}}] = 4y^{\mathcal{H}}_Q\) if \(m^\star = \mathcal{H}\) and \(\hat{Y}(m^\star,m^\star,Q) = 4\mathbb{E}_Q[Y|e^{\mathcal{L}}]= 4y^{\mathcal{L}}_Q\) if \(m^\star = \mathcal{L}\). Note that problem \ref{maxperiod1} is the same under myopia or unawareness of disagreement, because (i) there is no initial disagreement, and (ii) \(\delta^i V^i_{s_2}(m^\star,m^\star)\) drops out of the maximization. Hence \(\hat{Y}_u(m^\star,m^\star)=\hat{Y}_o(m^\star,m^\star) = \hat{Y}(m^\star,m^\star)\).  

Next, consider part (ii). Let players be unaware of disagreement. Fix \(i\in\{A,B\}\) and \(m_1^i\in M\). By unawareness of disagreement, player \(i\) plays as if the initial models of the two players were \((m^i_1,m^i_1)\). Since the two hypotheses compared by player \(i\) at the end of \(t=1\) coincide --- the alternative model is the same as the null model --- it has \(m^i_1 = m^i_2\) with certainty. Hence, if \(m^i_1 = \mathcal{H}\) then, \(i\)'s behavior is pinned down by \ref{like-minded1}. If \(m^i_1 = \mathcal{L}\) then, \(i\)'s behavior is pinned down by \ref{like-minded2}. It follows that 
\begin{gather*}
    \hat{Y}_u(\mathcal{H},\mathcal{L},Q) = 2(y^{\mathcal{H}}_Q+y^{\mathcal{L}}_Q) \\
    \hat{Y}_u(\mathcal{L},\mathcal{L},Q) = 4y^{\mathcal{L}}_Q \\
    \hat{Y}_u(\mathcal{H},\mathcal{H},Q) = 4y^{\mathcal{H}}_Q
\end{gather*}
which proves the first statement of part (ii).

Consider the second statement of part (ii), which assumes that \(\delta^i = 0\) for all \(i\in \{A,B\}\). Fix \(i\in \{A,B\}\). In equilibrium, for each \(m^A,m^B\in M\) and \(t=1,2\), \(s^i_{te}(m^A,m^B)\) must solve 
\begin{equation*}
 \max_{e\in \mathcal{E}} \mathbb{E}_{m^i}[u(Y,e)|e]   
\end{equation*}
so that \(s^i_{te}(m^A,m^B) = e^{m^i}.\) Consider now the probability of model change between the two periods. If \(m^i_1\ne m^{-i}_1\), player \(i\) could switch to model \(m^{-i}\) with positive probability between the two periods. 

Let \(Q = \mathcal{H}\) and \((m^A_1,m^B_1) = (\mathcal{H},\mathcal{L})\) at the beginning of period \(1\). Given the (myopic) equilibrium strategies, we have that \((e^A_1,e^B_1) = (e^\mathcal{H},e^\mathcal{L})\).  By definition of the test rule \ref{LR}, the likelihood that Ann switches to \(\mathcal{L}\) after period \(t=1\) is \(\phi^{\mathcal{H}}_\mathcal{L}(e^\mathcal{H}, e^{\mathcal{L}}|\mathcal{H}) = \alpha\), while the likelihood that Bob switches to \(\mathcal{H}\) after period \(t=1\) is \(\phi^{\mathcal{L}}_\mathcal{H}(e^\mathcal{H}, e^{\mathcal{L}}|\mathcal{H})\). 

We now show that  \(\phi^{\mathcal{L}}_\mathcal{H}(e^{\mathcal{H}}, e^{\mathcal{L}}|\mathcal{H}) \ge \alpha\). Consider the experiment \(\Pi_{(0,0)}\) arising if players chose \(e^A_1 = e^B_1 = 0\) in \(t=1\). If \(\mathcal{H}(\cdot|0) = \mathcal{L}(\cdot|0)\), \(\Pi_{(0,0)}\) is completely uninformative, because no output realization is more likely to be drawn from \(\mathcal{H}(\cdot|0)\) than from \(\mathcal{L}(\cdot|0)\). Under such an uninformative experiment, a switch occurs with probability \(\alpha\), independent of output realizations. Hence, it must be that \(\phi^{\mathcal{L}}_\mathcal{H}(0, 0|\mathcal{H}) = \alpha\). If instead \(\mathcal{H}(\cdot|0) \ne \mathcal{L}(\cdot|0)\), \(\Pi_{(0,0)}\) is not uninformative, so that, by lemma \ref{blackwell}, it must be \(\phi^{\mathcal{L}}_\mathcal{H}(0, 0|\mathcal{H}) \ge \alpha\), because \(\alpha\) is the power of the size-\(\alpha\) likelihood ratio test based on an uninformative experiment.  Hence \(\phi^{\mathcal{L}}_\mathcal{H}(0, 0|\mathcal{H}) \ge \alpha\). But then  \(\phi^{\mathcal{L}}_\mathcal{H}(e^{\mathcal{H}}, e^{\mathcal{L}}|\mathcal{H}) \ge \alpha\) follows by lemma \ref{lemmapower} and \(e^\mathcal{H}>0\).

Recall that if \(m^A_1 = \mathcal{H}\), \(m^B_1 = \mathcal{L}\), then \(e^A_1 = e^\mathcal{H}\) and \(e^B_1 = e^\mathcal{L}\). Hence, if \(Q = \mathcal{H}\), players' expected \textit{second-period} output conditional on first period models and actions, \(\mathbf{m}_1 = (\mathcal{H},\mathcal{L})\) and \(\mathbf{e}_1=(e^{\mathcal{H}}, e^{\mathcal{L}})\), satisfies

\begin{align*}
\mathbb{E}_{\mathcal{H}}[Y^A_2| m^A_1,m^B_1, e^{\mathcal{H}}, e^{\mathcal{L}}] &= \alpha y^\mathcal{L}_{\mathcal{H}} + (1-\alpha)y^\mathcal{H}_{\mathcal{H}} \\
\mathbb{E}_{\mathcal{H}}[Y^B_2| m^A_1,m^B_1, e^{\mathcal{H}}, e^{\mathcal{L}}] &= \phi^{\mathcal{L}}_\mathcal{H}(e^{\mathcal{H}}, e^{\mathcal{L}}|\mathcal{H}) y^\mathcal{H}_{\mathcal{H}} + (1-\phi^{\mathcal{L}}_\mathcal{H}(e^{\mathcal{H}}, e^{\mathcal{L}}|\mathcal{H}))y^\mathcal{L}_{\mathcal{H}}
\end{align*}
so that 
\begin{align}
\hat{Y}_o(\mathcal{H},\mathcal{L},\mathcal{H}) &= y^{\mathcal{H}}_{\mathcal{H}} + y^{\mathcal{L}}_{\mathcal{H}} + \mathbb{E}_{\mathcal{H}}[Y^A_2| m^A_1,m^B_1, e^{\mathcal{H}}, e^{\mathcal{L}}] + \mathbb{E}_{\mathcal{H}}[Y^B_2| m^A_1,m^B_1, e^{\mathcal{H}}, e^{\mathcal{L}}] \notag \\
& = 2y^{\mathcal{H}}_{\mathcal{H}} + 2y^{\mathcal{L}}_{\mathcal{H}} + (\phi^{\mathcal{L}}_\mathcal{H}(e^{\mathcal{H}}, e^{\mathcal{L}}|\mathcal{H}) - \alpha)(y^{\mathcal{H}}_{\mathcal{H}} - y^{\mathcal{L}}_{\mathcal{H}}) \notag \\
& \ge 2y^{\mathcal{H}}_{\mathcal{H}} + 2y^{\mathcal{L}}_{\mathcal{H}} = \frac{1}{2}\left(\hat{Y}_o(\mathcal{H},\mathcal{H},\mathcal{H})+\hat{Y}_o(\mathcal{L},\mathcal{L},\mathcal{H})\right). \label{compositionH}
\end{align}

By following analogous steps, one can show that if \(Q = \mathcal{L}\) it has 
so that 
\begin{align}
\hat{Y}_o(\mathcal{H},\mathcal{L},\mathcal{L}) &= 2y^{\mathcal{H}}_{\mathcal{L}} + 2y^{\mathcal{L}}_{\mathcal{L}} + (\phi^{\mathcal{H}}_\mathcal{L}(e^{\mathcal{H}}, e^{\mathcal{L}}|\mathcal{L}) - \alpha)(y^{\mathcal{L}}_{\mathcal{L}} - y^{\mathcal{H}}_{\mathcal{L}}) \notag \\
& \le 2y^{\mathcal{H}}_{\mathcal{L}} + 2y^{\mathcal{L}}_{\mathcal{L}} = \frac{1}{2} \left(\hat{Y}_o(\mathcal{H},\mathcal{H},\mathcal{L})+\hat{Y}_o(\mathcal{L},\mathcal{L},\mathcal{L})\right). \label{compositionL}
\end{align}
This proves the second set of inequalities of part (ii). Note that if \(\phi^{m^{\prime}}_m(\cdot,\cdot|m)\) is strictly increasing in its arguments it holds \(\phi^{m^{\prime}}_m(e^{\mathcal{H}},e^{\mathcal{L}}|m)>\alpha\). In such case, inequality \ref{compositionH} holds strictly, because \(y^{\mathcal{H}}_{\mathcal{H}}>y^{\mathcal{L}}_{\mathcal{H}}\) as noted in the proof of part (i).

Finally, for part (iii), it follows from the previous analysis that in any equilibrium of the game where players are myopic or unaware of disagreement, player \(i = A,B\) is expected to produce either  \(y^{\mathcal{H}}_Q\) or \(y^{\mathcal{L}}_Q\). Note that \(e^{\mathcal{H}}>e^{\mathcal{L}}\) and, using assumption \ref{ass1}, \(y^{\mathcal{H}}_Q>y^{\mathcal{L}}_Q\), so that the maximum expected output for the team is \(4y^{\mathcal{H}}_Q\). It is immediate to see that \(\hat{Y}(m^A,m^B, Q) = 4y^{\mathcal{H}}_Q \iff (m^A,m^B) = (\mathcal{H},\mathcal{H})\), which concludes the proof.

\subsection{Other Proofs}
\paragraph{Proof of Lemma \ref{lemmaeffort}} Consider any equilibrium profile \(s\in\hat{S}\) and fix \(i\in\{A,B\}\). Without loss of generality, let \(i = A\). In period \(t=2\), player \(A\) with model \(m^A_2\in M\) must choose the effort strategy solving \ref{maxperiod2}, so that it must hold 
\begin{equation*}
s^A_{2e}(\mathcal{H},m^B_2) = e^\mathcal{H},\quad s^A_{2e}(\mathcal{L},m^B_2) = e^\mathcal{L}, \quad\forall m^B_2\in M
\end{equation*}
Analogously, considering player \(B\), it must be
\begin{gather*}
s^B_{2e}(m^A_2,\mathcal{H}) = e^\mathcal{H},\quad s^B_{2e}(m^A_2, \mathcal{L}) = e^\mathcal{L}, \quad\forall m^A_2\in M.
\end{gather*}
Hence, the value of \(A\)'s expected second-period payoff when players play optimally give their models \((m^A_2,m^B_2)\) and \(B\) holds model \(m^B_2\in M\) in period \(t=2\) is \(\mathbb{E}_{m^A_2}[u(Y^A_2,e^{m^A_2})|e^{m^A_2}] + \mathbb{E}_{m^A_2}[v(Y^B_2)|e^{m^B_2}]\),
where \(e^{m^i_t}\) denotes \(e^{\mathcal{H}}\) if \(m^i_t = \mathcal{H}\) and \(e^{\mathcal{L}}\) if \(m^i_t = \mathcal{L}\). Consider any first-period model \(m^A_1\in M\). With expectations based on \(m^A_1\), it has
\begin{align*}
V^A_{s_{2e}, m^A_1}(m^A_2,m^B_2) &= \mathbb{E}_{m^A_1}[u(Y^A_2,e^{m^A_2})|e^{m^A_2}] + \mathbb{E}_{m^A_1}[v(Y^B_2)|e^{m^B_2}]. 
\end{align*}
If \(m^A_1 = m^B_1\), then, for all \(e^A_1, e^B_1\in\mathcal{E}\), 
\begin{equation}\label{Vlike}
\mathbb{E}_{m^A_1}[V^A_{s_{2e}, m^A_1}(m^A_2,m^B_2)|e^A_1,e^B_1,m^A_1,m^B_1] = \mathbb{E}_{m^A_1}[u(Y^A_2,e^{m^A_1})|e^{m^A_1}] + \mathbb{E}_{m^A_1}[v(Y^B_2)|e^{m^B_1}] 
\end{equation}
which does not depend on \(e^A_1, e^B_1\). If instead \(m^A_1 \ne m^B_1\)
then, for all \(e^A_1, e^B_1\in\mathcal{E}\),
\begin{align}\label{Vdis}
\mathbb{E}_{m^A_1}[V^A_{s_{2e}, m^A_1}(m^A_2,m^B_2)|e^A_1,e^B_1,m^A_1,m^B_1] = \alpha\mathbb{E}_{m^A_1}[u(Y^A_2,e^{m^B_1})|e^{m^B_1}]  + \notag \\
+(1-\alpha)\mathbb{E}_{m^A_1}[u(Y^A_2,e^{m^A_1})|e^{m^A_1}]  + \notag \\
+ \underbrace{\phi^{m^B_1}_{m^A_1}(e^A_1,e^B_1|m^A_1)\mathbb{E}_{m^A_1}[v(Y^B_2)|e^{m^A_1}] + (1-\phi^{m^B_1}_{m^A_1}(e^A_1,e^B_1|m^A_1))\mathbb{E}_{m^A_1}[v(Y^B_2)|e^{m^B_1}]}_{\mathbb{E}_{m^A_1}[v(Y^B_2)|e^{m^B_1}] + \phi^{m^B_1}_{m^A_1}(e^A_1,e^B_1|m^A_1)\left(\mathbb{E}_{m^A_1}[v(Y^B_2)|e^{m^A_1}] - \mathbb{E}_{m^A_1}[v(Y^B_2)|e^{m^B_1}]\right)}.
\end{align} 
which depends on \(e^A_1\) and \(e^B_1\) only through the term
\begin{equation}\label{mainforce}
\phi^{m^B_1}_{m^A_1}(e^A_1,e^B_1|m^A_1)\underbrace{\left(\mathbb{E}_{m^A_1}[v(Y^B_2)|e^{m^A_1}] - \mathbb{E}_{m^A_1}[v(Y^B_2)|e^{m^B_1}]\right)}_{\Delta_{m^A_1}(m^A_1,m^B_1)}.
\end{equation}
Let \(s_{1e}\) be the profile of equilibrium effort rules for period \(t=1\). It follows from \ref{Vlike}, \ref{Vdis} and \ref{mainforce} that we can write player \(A\)'s maximization problem in period \(t=1\) as 
\begin{equation}\label{maxAnn}
\max_{e\in \mathcal{E}} \left\{\mathbb{E}_{m^A_1}[u(Y,e)|e] + \mathbbm{1}_{(m_1^A\ne m_1^B)}\phi^{m^B_1}_{m^A_1}(e,s^B_{1e}(m^A_1,m^B_1)|m^A_1)\Delta_{m^A_1}(m^A_1,m^B_1)\right\}.
\end{equation}
Following the same steps for Bob,
\begin{equation}\label{maxBob}
\max_{e\in \mathcal{E}} \left\{\mathbb{E}_{m^B_1}[u(Y,e)|e] + \mathbbm{1}_{(m_1^A\ne m_1^B)}\phi^{m^A_1}_{m^B_1}(s^A_{1e}(m^A_1,m^B_1),e|m^B_1)\Delta_{m^B_1}(m^B_1,m^A_1)\right\}
\end{equation}
where \(\Delta_{m^B_1}(m^B_1,m^A_1) = \mathbb{E}_{m^B_1}[v(Y^A_2)|e^{m^B_1}] - \mathbb{E}_{m^B_1}[v(Y^A_2)|e^{m^A_1}]\). 

Since \(e^\mathcal{H}> e^\mathcal{L}\), by assumption \ref{ass1} and assumption \ref{ass2} it holds
\(\Delta_{\mathcal{H}}(\mathcal{H},\mathcal{L})> 0\ge \Delta_{\mathcal{L}}(\mathcal{L},\mathcal{H})\).

Ann and Bob period \(1\)'s effort rules given \(m^A_1,m^B_{1}\in M\) must solve \ref{maxAnn} and \ref{maxBob}, respectively. Assume that \(m^A_1 = \mathcal{H}\) and \(m^B_1 = \mathcal{L}\). We want to show that \(s^A_{1e}(\mathcal{H},\mathcal{L})\ge e^\mathcal{H}\) and \(s^B_{1e}(\mathcal{H},\mathcal{L})\le e^\mathcal{L}\). Consider any \(e<e^{\mathcal{H}}\). By assumption \ref{ass2}, \(e^{\mathcal{H}}\) is the unique maximizer of 
\(\mathbb{E}_{\mathcal{H}}[u(Y,e)|e]\). Hence, 
\begin{gather*}
\mathbb{E}_{\mathcal{H}}[u(Y,e^{\mathcal{H}})]> \mathbb{E}_{\mathcal{H}}[u(Y,e)|e] \\
\implies \mathbb{E}_{\mathcal{H}}[u(Y,e^{\mathcal{H}})] + \phi^{\mathcal{L}}_{\mathcal{H}}(e^{\mathcal{H}},s^B_{1e}(\mathcal{H},\mathcal{L})|\mathcal{H})\Delta_{\mathcal{H}}(\mathcal{H},\mathcal{L}) > \\
\mathbb{E}_{\mathcal{H}}[u(Y,e)|e] + \phi^{\mathcal{L}}_{\mathcal{H}}(e,s^B_{1e}(\mathcal{H},\mathcal{L})|\mathcal{H})\Delta_{\mathcal{H}}(\mathcal{H},\mathcal{L}),
\end{gather*}
which follows from lemma \ref{lemmapower} and \(\Delta_{\mathcal{H}}(\mathcal{H},\mathcal{L})> 0\). It follows that \(s^A_{1e}(\mathcal{H},\mathcal{L})\ge e^\mathcal{H}\). If \(\phi^{m^i}_{m^{-i}}(\cdot,\cdot|m^i)\) is strictly increasing in effort choices, note that \(e^\mathcal{H}>e^\mathcal{L}\) implies \(\Delta_{\mathcal{H}}(\mathcal{H},\mathcal{L})>0\) by assumption \ref{ass2}. Hence 
  \(\frac{\partial\phi^{\mathcal{H}}_{\mathcal{L}}(e^A,s_{1e}^B(\mathcal{H},\mathcal{L})|\mathcal{H})}{\partial e^A}\Delta_{\mathcal{H}}(\mathcal{H},\mathcal{L})>0\)
which implies that \(s_{1e}^A(\mathcal{H},\mathcal{L})>e^{\mathcal{H}}\). 

Let us now turn to Bob's equilibrium effort. Let \(e>e^{\mathcal{L}}\). Since by definition \(e^{\mathcal{L}}\) is the unique maximizer of 
\(\mathbb{E}_{\mathcal{L}}[u(Y,e)|e]\),
it must be that 
\begin{gather*}
\mathbb{E}_{\mathcal{L}}[u(Y,e^{\mathcal{L}})]> \mathbb{E}_{\mathcal{L}}[u(Y,e)|e] \\
\implies \mathbb{E}_{\mathcal{L}}[u(Y,e^{\mathcal{L}})] + \phi^{\mathcal{H}}_{\mathcal{L}}(s^A_{1e}(\mathcal{H},\mathcal{L}), e^{\mathcal{L}}|\mathcal{L})\Delta_{\mathcal{L}}(\mathcal{L},\mathcal{H}) > \\
\mathbb{E}_{\mathcal{L}}[u(Y,e)|e] + \phi^{\mathcal{H}}_{\mathcal{L}}(s^A_{1e}(\mathcal{H},\mathcal{L}),e|\mathcal{L})\Delta_{\mathcal{L}}(\mathcal{L},\mathcal{H}),
\end{gather*}
by lemma \ref{lemmapower} and \(\Delta_{\mathcal{H}}(\mathcal{H},\mathcal{L})\le 0\). It follows that \(s^B_{1e}(\mathcal{H},\mathcal{L})\le e^\mathcal{L}\). 

\paragraph{Proof of Propositon \ref{prop2formation}} Assume that \(\mathcal{L}\) discourages effort, i.e., \(e^\mathcal{L}=0\). By lemma \ref{lemmaeffort} we have \(s^A_{1e}(\mathcal{H},\mathcal{L})\ge e^\mathcal{H}\), and we can write \(s^A_{1e}(\mathcal{H},\mathcal{L}) = e^{\mathcal{H}} + \Delta_{e}\), for \(\Delta_{e} \equiv s^A_{1e}(\mathcal{H},\mathcal{L}) - e^{\mathcal{H}}\ge 0\). By lemma \ref{lemmaeffort}, it also follows \(s^B_{1e}(\mathcal{H},\mathcal{L}) \le s^B_{1e}(\mathcal{L},\mathcal{L})\) which implies \(s^B_{1e}(\mathcal{H},\mathcal{L}) = 0\). Following the same steps as for deriving \ref{compositionH} and \ref{compositionL}, but with effort levels \(s^A_{1e}(\mathcal{H},\mathcal{L})\) and \(s^B_{1e}(\mathcal{H},\mathcal{L})\) in the first period, we have that when \(Q =\mathcal{H}\)
\begin{align}
\hat{Y}(\mathcal{H},\mathcal{L},\mathcal{H}) &=  y^{\mathcal{H}}_{\mathcal{H}} + 2y^{\mathcal{L}}_{\mathcal{H}} + \mathbb{E}_\mathcal{H}[Y|e^{\mathcal{H}} + \Delta_{e}] + (\phi^{\mathcal{L}}_\mathcal{H}(e^{\mathcal{H}} + \Delta_e, e^{\mathcal{L}}|\mathcal{H}) - \alpha)(y^{\mathcal{H}}_{\mathcal{H}} - y^{\mathcal{L}}_{\mathcal{H}}) \notag \\
&\ge  2y^{\mathcal{H}}_{\mathcal{H}} + 2y^{\mathcal{L}}_{\mathcal{H}} + (\phi^{\mathcal{L}}_\mathcal{H}(e^{\mathcal{H}}, e^{\mathcal{L}}|\mathcal{H}) - \alpha)(y^{\mathcal{H}}_{\mathcal{H}} - y^{\mathcal{L}}_{\mathcal{H}}) =\hat{Y}_o(\mathcal{H},\mathcal{L},\mathcal{H}) \label{composition_disH}
\end{align}
where the inequality follows from \(\Delta_e\ge0\), FOSD-monotonicity and lemma \ref{lemmapower}. By proposition \ref{prop2formation} it holds that \(2\hat{Y}_o(\mathcal{H},\mathcal{L},\mathcal{H}) \ge \hat{Y}_o(\mathcal{H},\mathcal{H},\mathcal{H}) + \hat{Y}_o(\mathcal{L},\mathcal{L},\mathcal{H})\) and also that, for \(m\in \{\mathcal{H},\mathcal{L}\}\), it has \(\hat{Y}_o(m,m, Q) = \hat{Y}(m,m, Q)\). Hence
\begin{equation}\label{equiv}
2\hat{Y}_o(\mathcal{H},\mathcal{L},\mathcal{H})\ge \hat{Y}(\mathcal{H},\mathcal{H},\mathcal{H}) + \hat{Y}(\mathcal{L},\mathcal{L},\mathcal{H}).
\end{equation}
Combining \ref{composition_disH} and \ref{equiv}, we obtain
\begin{equation}\label{prop2part1}
2\hat{Y}(\mathcal{H},\mathcal{L},\mathcal{H})\ge 2\hat{Y}_o(\mathcal{H},\mathcal{L},\mathcal{H})\ge \hat{Y}(\mathcal{H},\mathcal{H},\mathcal{H}) + \hat{Y}(\mathcal{L},\mathcal{L},\mathcal{H}).
\end{equation}
By the same arguments of the previous proof, when \(\phi^{\mathcal{L}}_\mathcal{H}(\cdot, \cdot|\mathcal{H})\) is strictly increasing in \(e^A\), we have \(\Delta_e>0\), which implies that \ref{prop2part1} holds strictly. To see why the ranking is ambiguous when \(Q = \mathcal{L}\), note that by following analogous steps as for \(Q = \mathcal{H}\), one obtains,
\begin{align}
\hat{Y}(\mathcal{H},\mathcal{L},\mathcal{L}) &=  y^{\mathcal{H}}_{\mathcal{L}} + 2y^{\mathcal{L}}_{\mathcal{L}} + \mathbb{E}_\mathcal{L}[Y|e^{\mathcal{H}} + \Delta_{e}] + (\phi^{\mathcal{H}}_\mathcal{L}(e^{\mathcal{H}} + \Delta_e, e^{\mathcal{L}}|\mathcal{L}) - \alpha)(y^{\mathcal{L}}_{\mathcal{L}} - y^{\mathcal{H}}_{\mathcal{L}}) \notag \\
&\gtreqless  2y^{\mathcal{H}}_{\mathcal{L}} + 2y^{\mathcal{L}}_{\mathcal{L}} =\frac{1}{2}\left(\hat{Y}(\mathcal{H},\mathcal{H},\mathcal{L}) + \hat{Y}(\mathcal{L},\mathcal{L},\mathcal{L})\right).\label{composition_disL}
\end{align}
In particular, from \ref{composition_disL} one obtains that 
\begin{align*}
2\hat{Y}(\mathcal{H},\mathcal{L},\mathcal{L})&> \hat{Y}(\mathcal{H},\mathcal{H},\mathcal{L}) + \hat{Y}(\mathcal{L},\mathcal{L},\mathcal{L}) \\
\iff \mathbb{E}_\mathcal{L}[Y|e^{\mathcal{H}} + \Delta_{e}] - y^\mathcal{H}_{\mathcal{L}}&> \left(\phi^{\mathcal{H}}_\mathcal{L}(e^{\mathcal{H}} + \Delta_e, e^{\mathcal{L}}|\mathcal{L}) - \alpha\right)(y^{\mathcal{H}}_{\mathcal{L}} - y^{\mathcal{L}}_{\mathcal{L}}),
\end{align*} 
which is a condition for the disagreeing team to outperform the average like-minded team.

\paragraph{Proof of Proposition \ref{prop3}} See example \ref{ex3} for the first part. To see why the condition is sufficient for \(\hat{Y}(\mathcal{H},\mathcal{L},Q)> \hat{Y}(\mathcal{H},\mathcal{H},Q)\), note that it must be \(s^B_{1e}(\mathcal{H},\mathcal{L})\le e^{\mathcal{L}}\), by lemma \(\ref{lemmaeffort}\). Hence the condition implies
\begin{equation*}
\frac{\partial}{\partial e^A_1}\mathbb{E}_{\mathcal{H}}[u(Y^A,e^A_1)|e^A_1] + \frac{\partial}{\partial e^A_1}\phi_{\mathcal{H}}^{\mathcal{L}}(e^A_1,s^B_{1e}(\mathcal{H},\mathcal{L})|\mathcal{H})\Delta>0,
\end{equation*}
for any \(e^A_1\in [e^{\mathcal{H}},\hat{e}_Q]\). Hence \(s^B_{1e}(\mathcal{H},\mathcal{L}) > \hat{e}_Q\) and \(\hat{Y}(\mathcal{H},\mathcal{L})>\hat{Y}(\mathcal{H},\mathcal{H})\), by assumption \ref{ass1}.

\paragraph{Proof of Proposition \ref{prop4fixed}} Let \(K = \{x,y\}\), and let \(m^A_2 = (m^A_{2x},m^A_{2y})\in M\) and \(m^B_2 = (m^B_{2x},m^B_{2y})\in M\) be Ann and Bob's period \(t=2\) models. The expected utility that \(i=A,B\) obtains by optimally operating \(k\in \{x,y\}\) in period \(t=2\)
is \(\mathbb{E}_{m^i_{2k}}[u(Y,e^{m^i_{2k}})|e^{m^i_{2k}},k]\) 
where \(e^{\mathcal{H}_k} = e^{\mathcal{H}}\), \(e^{\mathcal{L}_k} = e^{\mathcal{L}}\). By assumption \ref{ass2},
\(
   \mathbb{E}_{\mathcal{H}_{k}}[u(Y,e^{\mathcal{H}_{k}})|e^{\mathcal{H}_{k}},k]> \mathbb{E}_{\mathcal{L}_{k}}[u(Y,e^{\mathcal{L}_{k}} )|e^{\mathcal{L}_{k}},k].
\)
In equilibrium, the technology choice rule will therefore satisfy
\begin{equation*}
s^i_{2k}(m^A_2,m^B_2) \in \arg\max_{k\in \{x, y\}} m^i_{2k}
\end{equation*}
where we set \(\mathcal{H}>\mathcal{L}\) by convention. Let \(k^\star = s^i_{2k}(m^A_2,m^B_2)\). The effort rule \(s^i_{2e}\) will satisfy
\begin{gather*}
s^i_{2e}(m^A_2,m^B_2) = e^{\mathcal{H}} \iff m^i_{2k^\star} = \mathcal{H} \\
s^i_{2e}(m^A_2,m^B_2) = e^{\mathcal{L}} \iff m^i_{2k^\star} = \mathcal{L}
\end{gather*}

Consider now any two first-period models \(m^A_1,m^B_1\in M\). Based on \(m^A_1\), the value of the second period, expressed as a function of \((m^A_2,m^B_2)\in M^2\) is
\begin{equation*}
\resizebox{1\textwidth}{!}{$V^A_{s_{2}, m^A_1}(m^A_2,m^B_2) = \mathbb{E}_{m^A_1}[u(Y^A_2,s^A_{2e}(m^A_2,m^B_2))] |s^A_2(m^A_2,m^B_2)] + \mathbb{E}_{m^A_1}[v(Y^B_2)|s^B_2(m^A_2,m^B_2)].$} 
\end{equation*}
Given that \(i\)'s second period choices depend only on \(m^i_2\), for \(e^A_1,e^B_1\in \mathcal{E}\) and \(k^A_1,k^B_1\in \mathcal{E}\),
\begin{footnotesize}
\begin{align*}
\mathbb{E}_{m^A_1}[V^A_{s_{2}, m^A_1}(m^A_2,m^B_2)|&e^A_1,k^A_1,e^B_1,k^B_1,m^A_1,m^B_1]= \alpha\mathbb{E}_{m^A_1}[u(Y^A_2,s^A_{2e}(m^B_1,m^B_1)) |s^A_2(m^B_1,m^B_1)] \\
& + (1-\alpha)\mathbb{E}_{m^A_1}[u(Y^A_2,s^A_{2e}(m^A_1,m^B_1)) |s^A_2(m^A_1,m^B_1)] \\
& + \phi^{m^B_1}_{m^A_1}(e^A_1,e^B_1,k^A_1,k^B_1|m^A_1)\mathbb{E}_{m^A_1}[v(Y^B_2)|s^B_2(m^A_1,m^A_1)] \\ 
& + (1-\phi^{m^B_1}_{m^A_1}(e^A_1,e^B_1,k^A_1,k^B_1|m^A_1))\mathbb{E}_{m^A_1}[v(Y^B_2)|s^B_2(m^A_1,m^B_1)],
\end{align*}
\end{footnotesize}
which depends on \((e^A_1,e^B_1,k^A_1,k^B_1)\) only through \(\phi^{m^B_1}_{m^A_1}\). Hence, for fixed \(\hat{k}_1^A, \hat{k}_1^B \in K\) and Bob's equilibrium effort \(s^B_{1e}(m^A_1,m^B_1)\), Ann's effort \(s^A_{1e}(m^A_1,m^B_1)\) must be a solution to
\begin{equation}\label{maxAnn2}
\resizebox{.99\textwidth}{!}{$\max_{e\in \mathcal{E}} \left\{\mathbb{E}_{m^A_1}[u(Y,e)|e, \hat{k}^A_1] + \mathbbm{1}_{(m_1^A\ne m_1^B)}\phi^{m^B_1}_{m^A_1}(e,s^B_{1e}(m^A_1,m^B_1), \hat{k}^A_1,\hat{k}^B_1|m^A_1)\Delta_{m^A_1}(m^A_1,m^B_1)\right\}$}
\end{equation}
where \(\Delta_{m^A_1}(m^A_1,m^B_1) = \mathbb{E}_{m^A_1}[v(Y^B_2)|s^B_2(m^A_1,m^A_1)]- \mathbb{E}_{m^A_1}[v(Y^B_2)|s^B_2(m^A_1,m^B_1)]\). 

Similarly, Bob's effort rule \(s^B_1(m^A_1,m^B_1)\) must be a solution to
\begin{equation}\label{maxBob2}
\resizebox{.99\textwidth}{!}{$\max_{e\in \mathcal{E}} \left\{\mathbb{E}_{m^B_1}[u(Y,e)|e,\hat{k}^B_1] + \mathbbm{1}_{(m_1^A\ne m_1^B)}\phi^{m^A_1}_{m^B_1}(s^A_{1e}(m^A_1,m^B_1),e, \hat{k}^A_1,\hat{k}^B_1|m^B_1)\Delta_{m^B_1}(m^B_1,m^A_1)\right\}$}
\end{equation}
where \(\Delta_{m^B_1}(m^B_1,m^A_1) = \mathbb{E}_{m^B_1}[v(Y^A_2)|s^A_2(m^B_1,m^B_1)]- \mathbb{E}_{m^B_1}[v(Y^A_2)|s^A_2(m^A_1,m^B_1)]\).

Now assume that the true technology is such that \(Q_x = Q_y = \hat{Q}\). Consider like-minded teams first. Inspecting the above maximization problems, it is easily seen that
\(s^i_{1e}((\mathcal{H}_x,\mathcal{H}_y),(\mathcal{H}_x,\mathcal{H}_y)) = e^{\mathcal{H}}\) and \( s^i_{1e}((\mathcal{L}_x,\mathcal{L}_y),(\mathcal{L}_x,\mathcal{L}_y)) = e^{\mathcal{L}}\). If Ann and Bob share model \((\mathcal{H}_x,\mathcal{L}_y)\), each of them will exert effort \(e^\mathcal{H}\) if initially assigned to \(x\) and \(e^\mathcal{L}\) if assigned to \(y\). Similarly, if they share \((\mathcal{L}_x,\mathcal{H}_y)\), they will exert effort \(e^\mathcal{H}\) if assigned to \(y\) and \(e^\mathcal{L}\) if assigned to \(x\). Hence, for each \(i=A,B\) and \(\hat{k}^i_1\in\{x,y\}\), \(s^i_{1e}((\mathcal{H}_x,\mathcal{L}_y),(\mathcal{H}_x,\mathcal{L}_y)) \le e^{\mathcal{H}}\) and
\(s^i_{1e}((\mathcal{L}_x,\mathcal{H}_y),(\mathcal{L}_x,\mathcal{H}_y)) \le e^{\mathcal{H}}\). Because of initial like-mindedness, no model change can occur between period \(t=1\) and \(t=2\), so that effort profiles are equal across periods. Hence, if \(m^A_1=m^B_1\), it has \(\hat{Y}(\hat{k}^A_1,\hat{k}^B_1,m_1^A,m_1^B) \le 4y^{\mathcal{H}}_{\hat{Q}}\),
where \(y^m_{\hat{Q}} = \mathbb{E}_{Q_x}[Y|e^{m},x] = \mathbb{E}_{Q_y}[Y|e^{m},y]\), for \(m\in \{\mathcal{H},\mathcal{L}\}\). Now, consider the case where \(m^A_1 = (\mathcal{H}_x,\mathcal{L}_y)\) and \(m^B_1 = (\mathcal{L}_x,\mathcal{H}_y)\). We show that if \(\hat{k}^A_1 = x\), then 
\begin{equation} \label{effort4a}
    s^A_{1e}((\mathcal{H}_x,\mathcal{L}_y),(\mathcal{L}_x,\mathcal{H}_y))\ge e^{\mathcal{H}}
\end{equation}
and if \(\hat{k}^B_1 = y\) then
\begin{equation} \label{effort4b}
s^B_{1e}((\mathcal{H}_x,\mathcal{L}_y),(\mathcal{L}_x,\mathcal{H}_y))\ge e^{\mathcal{H}}.
\end{equation}
To see why inequalities \ref{effort4a} and \ref{effort4b} hold, it is useful to prove the following lemma.
\begin{lemma}\label{lemma4}
Let \(|K| = 2\). For each \(k^A,k^B\in K\) and \(i\in I\), \(\phi_{m^{i}}^{m^{-i}}(e^A,e^B,k^A,k^B|m^i)\) is increasing in \(e^A\) and \(e^B\).
\end{lemma}
Fix \(k^A, k^B\in K\) and \(m^A,m^B\in M\). First, note that if \(m^{A} = m^{B}\) the lemma holds trivially. Similarly, if \(k^A = k^B = k^\star\) and \(m^A_{k^\star} = m^B_{k^\star}\), for some \(k^\star \in K\), then only information about technology \(k^\star\) arrives during \(t=1\). Since players hold the same view for technology \(k^\star\) and observe the same information, they must use the exact same rejection rule. Hence \(\phi_{m^{-i}}^{m^{i}}(e^A,e^B,k^\star,k^\star|m^i) = \alpha\implies \phi_{m^{i}}^{m^{-i}}(e^A,e^B,k^\star,k^\star|m^i) = \alpha\), constant in \(e^A\) and \(e^B\). 

Consider now the case where there exists \(k\in\{k^A,k^B\}\) such that \(m^A_{k}\ne m^B_{k}\). Without loss of generality, assume \(k^A\) is such that \(m^A_{k^A}\ne m^B_{k^A}\). Fix effort \(e^A,e^B\in \mathcal{E}\). \(\Pi_{(e^A,k^A)} = (m^A_{k^A}(\cdot|e^A), m^B_{k^A}(\cdot|e^A), \mathcal{Y})\) is an informative experiment for states \(m^A\) and \(m^B\). There are two possible cases. First, if \(m^A_{k^B} = m^B_{k^B}\), then \((y^B,e^B)\) are uninformative when it comes to discriminating between \(m^A\) and  \(m^B\). Because \(y^A\) and \(y^B\) are independent, test \ref{LR} will only use information about \((y^A,e^A,k^A)\). Hence, we have \(\phi_{m^{A}}^{m^{B}}(e^A,e^B,k^A,k^B|m^A) = \phi_{m^{A}_{k^A}}^{m^{B}_{k^A}}(e^A,k^A|m^A_{k^A})\) which is increasing in \(e^A\) by lemma \ref{blackwell}. The second case is \(m^A_{k^B} \ne m^B_{k^B}\), in which case \(\Pi_{(e^B,k^B)} = (m^A_{k^B}(\cdot|e^B), m^B_{k^B}(\cdot|e^B), \mathcal{Y})\) is also an informative experiment with states \(m^A\) and \(m^B\). Assume that this is the case and define the composite experiment \(\Pi_{(e^A,k^A,e^B,k^B)} = (\Pi_{(e^A,k^A)},\Pi_{(e^B,k^B)})\). Next, consider the composite experiment \(\Pi_{(\hat{e}^A,k^A,e^B,k^B)} = (\Pi_{(\hat{e}^A,k^A)},\Pi_{(e^B,k^B)})\), where \(\hat{e}^A>e^A\). By assumption \ref{ass3}, \(\Pi_{(\hat{e}^A,k^A)}\) is more informative than \(\Pi_{(e^A,k^A)}\). Since \(\Pi_{(\hat{e}^A,k^A)}\), \(\Pi_{(e^A,k^A)}\) and \(\Pi_{(e^B,k^B)}\) are independent, \(\Pi_{(\hat{e}^A,k^A,e^B,k^B)}\) is more informative than \(\Pi_{(e^A,k^A,e^B,k^B)}\) --- the proof is analogous to the one provided for lemma \ref{lemmapower}. Hence, by lemma \ref{blackwell}, \(\phi_{m^{A}}^{m^{B}}(\hat{e}^A,e^B,k^A,k^B|m^A)\ge\phi_{m^{A}}^{m^{B}}(e^A,e^B,k^A,k^B|m^A)\), which proves that \(\phi_{m^{A}}^{m^{B}}(e^A,e^B,k^A,k^B|m^A)\) is increasing in \(e^A\). The proof that \(\phi_{m^{A}}^{m^{B}}(e^A,e^B,k^A,k^B|m^A)\) is increasing in \(e^B\) is a replication of the same argument. The proof that \(\phi_{m^{B}}^{m^{A}}(e^A,e^B,k^A,k^B|m^B)\) is increasing in \(e^A\) and \(e^B\) is analogous.

Let us go back to showing that \ref{effort4a} and \ref{effort4b} hold. Fix \(e<e^{\mathcal{H}}\). By definition and uniqueness of \(e^{\mathcal{H}}\), given \(m^A_1 = (\mathcal{H}_x,\mathcal{L}_y
)\) we have \(\mathbb{E}_{m^A_1}[u(Y,e^\mathcal{H})|x]> \mathbb{E}_{m^A_1}[u(Y,e)|x]\), hence
\begin{align*}
\mathbb{E}_{m^A_1}[u(Y,e^\mathcal{H})|x] + \phi^{m^B_1}_{m^A_1}\left(e^\mathcal{H},s^B_{1e}(m^A_1,m^B_1), x,\hat{k}^B_1|m^A_1\right)\Delta_{m^A_1}(m^A_1,m^B_1) > \\
>  \mathbb{E}_{m^A_1}[u(Y,e)|x] + \phi^{m^B_1}_{m^A_1}\left(e,s^B_{1e}(m^A_1,m^B_1), x,\hat{k}^B_1|m^A_1\right) \Delta_{m^A_1}(m^A_1,m^B_1),
\end{align*}
since it has \(\phi^{m^B_1}_{m^A_1}\left(e^\mathcal{H},s^B_{1e}(m^A_1,m^B_1), x,\hat{k}^B_1|m^A_1\right)\ge\phi^{m^B_1}_{m^A_1}\left(e,s^B_{1e}(m^A_1,m^B_1), x,\hat{k}^B_1|m^A_1\right)\) by lemma \ref{lemma4} and \(\Delta_{m^A_1}(m^A_1,m^B_1) > 0\), which follows from \(e^\mathcal{H}>0\) and the strict FOSD assumption \ref{ass2}. This proves that, if \(\hat{k}^A_1 = x\), \(s^A_{1e}((\mathcal{H}_x,\mathcal{L}_y),(\mathcal{L}_x,\mathcal{H}_y))\ge e^{\mathcal{H}}\). The proof for Bob when \(\hat{k}^B_1 =y\) follows analogous steps, so that, if  \(\hat{k}^B_1 = y\), then \(s^B_{1e}((\mathcal{H}_x,\mathcal{L}_y),(\mathcal{L}_x,\mathcal{H}_y))\ge e^{\mathcal{H}}\). 

Next, note that, if Ann and Bob's initial models are \((\mathcal{H}_x,\mathcal{L}_y)\) and \((\mathcal{L}_x,\mathcal{H}_y)\), at \(t=2\) both players will certainly have one view of type \(\mathcal{H}\) --- although they might still hold such view for different technologies. This means that, at \(t=2\), both players will choose effort \(e^\mathcal{H}\). Hence, by strict FOSD monotonicity and \(Q_x = Q_y\),
\begin{equation*}
\hat{Y}(x,y,(\mathcal{H}_x,\mathcal{L}_y),(\mathcal{L}_x,\mathcal{H}_y), Q) \ge 4y^{\mathcal{H}}_{\hat{Q}} \ge \hat{Y}(k^A,k^B,m,m, Q)
\end{equation*}
for each \(k^A,k^B\in K\) and \(m\in M\). By the same argument used in the previous proofs, the inequality holds strictly if \(\phi^{m^i_1}_{m^{-i}_1}\left(\cdot,\cdot, \hat{k}^A_1,\hat{k}^B_1|m^i_1\right)\) is strictly increasing in \(e^i\).

\paragraph{Proof of Proposition \ref{prop5end}} We start from part (i). When the choice of \(k^i_1\) is endogenous, problems \ref{maxAnn2} and \ref{maxBob2} become, respectively,
\begin{equation}
\resizebox{.99\textwidth}{!}{$\max_{e\in \mathcal{E},k\in K} \left\{\mathbb{E}_{m^A_1}[u(Y,e)|e,k] + \mathbbm{1}_{(m_1^A\ne m_1^B)}\phi^{m^B_1}_{m^A_1}(e,k, s^B_{1}(m^A_1,m^B_1)|m^A_1)\Delta_{m^A_1}(m^A_1,m^B_1)\right\},$}
\end{equation}
where \(\Delta_{m^A_1}(m^A_1,m^B_1) = \mathbb{E}_{m^A_1}[v(Y^B_2)|s^B_2(m^A_1,m^A_1)]- \mathbb{E}_{m^A_1}[v(Y^B_2)|s^B_2(m^A_1,m^B_1)]\). 

Similarly, for Bob it holds
\begin{equation}
\resizebox{.99\textwidth}{!}{$\max_{e\in \mathcal{E}, k\in K} \left\{\mathbb{E}_{m^B_1}[u(Y,e)|e,k] + \mathbbm{1}_{(m_1^A\ne m_1^B)}\phi^{m^A_1}_{m^B_1}(s^A_{1}(m^A_1,m^B_1),e, k|m^B_1)\Delta_{m^B_1}(m^B_1,m^A_1)\right\},$}
\end{equation}
with \(\Delta_{m^B_1}(m^B_1,m^A_1) = \mathbb{E}_{m^B_1}[v(Y^A_2)|s^A_2(m^B_1,m^B_1)]- \mathbb{E}_{m^B_1}[v(Y^A_2)|s^A_2(m^A_1,m^B_1)]\). 

It is sufficient so show that when \((m^A_1,m^B_1) = ((\mathcal{H}_x,\mathcal{L}_y),(\mathcal{L}_x,\mathcal{H}_y))\), then \(s^A_{1k}(m^A_1,m^B_1) = x\) and \(s^B_{1k}(m^A_1,m^B_1) = y\). Fix an arbitrary \(e^A\in\mathcal{E}\). If Ann exerts \(e^A\in \mathcal{E}\) in technology \(x\), experiment \(\Pi_{(e^A,x)}=(\mathcal{H}(\cdot|e^A),\mathcal{L}(\cdot|e^A),\mathcal{Y})\) is generated: \(Y^A\) is drawn from \(\mathcal{H}(\cdot|e^A)\) if \(Q = m_1^A\), while it is drawn from   \(\mathcal{L}(\cdot|e^A)\) if \(Q = m_1^B\). If Ann exerts \(e^A\in \mathcal{E}\) in technology \(y\), experiment \(\Pi_{(e^A,y)} = (\mathcal{L}(\cdot|e^A),\mathcal{H}(\cdot|e^A),\mathcal{Y})\) is generated: \(Y^A\) is drawn from \(\mathcal{L}(\cdot|e^A)\) if \(Q = m_1^A\), while it is drawn from   \(\mathcal{H}(\cdot|e^A)\) if \(Q = m_1^B\). The two experiments, \(\Pi_{(e^A,x)}\) and \(\Pi_{(e^A,y)}\) are Blackwell equivalent by equal falsifiability. Next, fix any \(e^B\in \mathcal{E}\) and \(k^B\in K\). When Bob exerts \(e^B\) with technology \(k^B\), the experiment \(\Pi_{(e^B,k^B)}\) is generated. Given the independence of \(\Pi_{(e^A,x)}\), \(\Pi_{(e^A,y)}\) and \(\Pi_{(e^B,k^B)}\), experiment \((\Pi_{(e^A,x)},\Pi_{(e^B,k^B)})\) is Blackwell equivalent to \((\Pi_{(e^A,y)},\Pi_{(e^B,k^B)})\), by an argument analogous to the one used in the proof of lemma \ref{lemmapower}.
It follows that, for any \(e^A,e^B\in \mathcal{E}\) and \(k^B\in K\), it holds \(\phi^{m^B_1}_{m^A_1}(e^A,e^B,x,k^B_1|m^A_1) = \phi^{m^B_1}_{m^A_1}(e^A,e^B,y,k^B_1|m^A_1)\) by lemma \ref{blackwell}. But Ann expects the effort to pay off more when invested in technology \(x\), because she starts the game with views \((\mathcal{H}_x,\mathcal{L}_y)\), hence  \(\mathbb{E}_{m^A_1}[u(Y,e)|e,x]>\mathbb{E}_{m^A_1}[u(Y,e)|e,y]\) for each \(e\in \mathcal{E}\). Hence she always strictly prefers to exert effort \(e^A\) in \(x\) rather than in \(y\), which implies \(s^A_{1k}((\mathcal{H}_x,\mathcal{L}_y),(\mathcal{L}_x,\mathcal{H}_y)) = x\). By a similar argument, \(s^B_{1k}((\mathcal{H}_x,\mathcal{L}_y),(\mathcal{L}_x,\mathcal{H}_y)) = y\). Given that Ann and Bob start producing using \(x\) and \(y\) respectively, the statement of part (i) follows from proposition \ref{prop4fixed}.

Next, we prove part (ii). Fix any equilibrium \(s\in\hat{S}\). Let \(m\in M\). By the standard arguments of the previous proposition, when players start the game agreeing on \(m\), in equilibrium they will exert at most effort \(e^{\mathcal{H}}\) in every period. Since models \((\mathcal{H},\mathcal{L})\) and \((\mathcal{L},\mathcal{H})\) are both correct with probability \(\frac{1}{2}\), and effort and technology choices are only contingent on models --- not changed between periods --- it holds \(\mathbb{E}_p[Y_s(m,m,Q)] \le 2(y^{\mathcal{H}}_{\mathcal{H}}+y^{\mathcal{H}}_{\mathcal{L}})\).

Now, consider the disagreeing team with \(\mathbf{m}_1 = ((\mathcal{H}_x,\mathcal{L}_y),(\mathcal{L}_x,\mathcal{H}_y))\). By our previous results, \(s\) must be such that, for \(i=A,B\), \(s^i_1((\mathcal{H}_x,\mathcal{L}_y), (\mathcal{L}_x,\mathcal{H}_y)) = e^{\mathcal{H}}+\hat{\Delta}_e\) for some  \(\hat{\Delta}_e\ge 0\).

If \(Q = (\mathcal{H}_x,\mathcal{L}_y)\), then it holds
\begin{gather}
Y_s(m^A_1, m^B_1,Q) = \mathbb{E}_{\mathcal{H}}[Y|e^{\mathcal{H}}+\hat{\Delta}_e,x] + \mathbb{E}_{\mathcal{L}}[Y|e^{\mathcal{H}}+\hat{\Delta}_e,y] + y^{\mathcal{H}}_{\mathcal{H}} \notag \\ 
+ y^{\mathcal{H}}_{\mathcal{L}} + (\phi^{m^A_1}_{m^B_1}(e^{\mathcal{H}}+\hat{\Delta}_e,e^{\mathcal{H}}+\hat{\Delta}_e,x,y|m^A_1)-\alpha)(y^{\mathcal{H}}_{\mathcal{H}} - y^{\mathcal{H}}_{\mathcal{L}}) \label{Y_high}
\end{gather}
If instead \(Q = (\mathcal{L}_x,\mathcal{H}_y)\), then it holds
\begin{gather}
Y_s(m^A_1, m^B_1,Q) = \mathbb{E}_{\mathcal{H}}[Y|e^{\mathcal{H}}+\hat{\Delta}_e,y] + \mathbb{E}_{\mathcal{L}}[Y|e^{\mathcal{H}}+\hat{\Delta}_e,x] + y^{\mathcal{H}}_{\mathcal{H}} \notag \\ 
+ y^{\mathcal{H}}_{\mathcal{L}} + (\phi^{m^B_1}_{m^A_1}(e^{\mathcal{H}}+\hat{\Delta}_e,e^{\mathcal{H}}+\hat{\Delta}_e,x,y|m^B_1)-\alpha)(y^{\mathcal{H}}_{\mathcal{H}} - y^{\mathcal{H}}_{\mathcal{L}}) \label{Y_low}
\end{gather}
Using \ref{Y_high}, \ref{Y_low}, and \(\mathbb{E}_{m}[Y|e^{\mathcal{H}}+\hat{\Delta}_e,x] = \mathbb{E}_{m}[Y|e^{\mathcal{H}}+\hat{\Delta}_e,y]\) for \(m\in\{\mathcal{H},\mathcal{L}\}\), we have
\begin{gather*}
\mathbb{E}_p[Y_s(m^A_1,m^B_1,Q)] = y^{\mathcal{H}}_{\mathcal{H}}+y^{\mathcal{H}}_{\mathcal{L}} +  \mathbb{E}_{\mathcal{H}}[Y|e^{\mathcal{H}}+\hat{\Delta}_e,x] + \mathbb{E}_{\mathcal{L}}[Y|e^{\mathcal{H}}+\hat{\Delta}_e,y] \\
+ (\phi^{m^A_1}_{m^B_1}(e^{\mathcal{H}}+\hat{\Delta}_e,e^{\mathcal{H}}+\hat{\Delta}_e,x,y|m^A_1)-\alpha)(y^{\mathcal{H}}_{\mathcal{H}} - y^{\mathcal{H}}_{\mathcal{L}}).
\end{gather*}
Hence 
\begin{gather*}
\mathbb{E}_p[Y_s(m^A_1,m^B_1,Q)] - \mathbb{E}_p[Y_s(m,m,Q)] \ge \underbrace{(\mathbb{E}_{\mathcal{H}}[Y|e^{\mathcal{H}}+\hat{\Delta}_e,x] - y^{\mathcal{H}}_{\mathcal{H}})}_{\ge 0} \\
+ \underbrace{(\mathbb{E}_{\mathcal{L}}[Y|e^{\mathcal{H}}+\hat{\Delta}_e,y] - y^{\mathcal{H}}_{\mathcal{L}})}_{\ge 0} \\
+\underbrace{\phi^{m^A_1}_{m^B_1}(e^{\mathcal{H}}+\hat{\Delta}_e,e^{\mathcal{H}}+\hat{\Delta}_e,x,y|m^A_1)-\alpha)(y^{\mathcal{H}}_{\mathcal{H}} - y^{\mathcal{H}}_{\mathcal{L}})}_{\ge 0} \ge 0,
\end{gather*}
where the first two terms on the right-hand side are non-negative by the FOSD-monotonicity part of assumption \ref{ass2}, while the last term is non-negative by an argument analogous to the one presented in proposition \ref{prop1bench} when dealing with the single technology case. In particular, if the experiment generated in the first period after players make choices \((e^A_1,e^B_1,k^A_1,k^B_1)\) was completely uninformative about \(Q\) we would have \(\phi^{m^A_1}_{m^B_1}(e^A_1,e^B_1,k^A_1,k^B_1|m^A_1)=\alpha\) so that, by lemma \ref{lemma4}, it must be \(\phi^{m^A_1}_{m^B_1}(e^{\mathcal{H}}+\hat{\Delta}_e,e^{\mathcal{H}}+\hat{\Delta}_e,x,y|m^A_1)\ge \alpha\). Note that the first two terms capture the production gain due to the additional persuasion effort in period \(t=1\), while the last term captures the benefit of the additional information produced, which pushes both players to produce with the best technology in period \(t=2\). Finally, if \(\phi^{m^{-i}_1}_{m^i_1}(\cdot,\cdot,k^A_1,k^B_1|m^A_1)\) is strictly increasing in \(e^A_1\) and \(e^B_1\) then \(\hat{\Delta}_e>0\), in which case the strict FOSD-monotonicity assumption for \(\mathcal{H}\) implies \(\mathbb{E}_p[\hat{Y}(m^A_1,m^B_1,Q)] - \mathbb{E}_p[\hat{Y}(m,m,Q)]>0\).
\end{document}